\documentclass[11pt]{article}
\usepackage[dvips]{graphicx,epsfig}
\usepackage{bm}
\usepackage[normalem]{ulem}
\usepackage{amsmath,amssymb}
\usepackage{dsfont}
\usepackage{braket}
\usepackage[nosort]{cite}
\usepackage{color}
\usepackage{slashed}
\usepackage[T1]{fontenc}
\usepackage{hyphenat}
\usepackage{amsfonts,mathrsfs,mathdots}
\usepackage{multicol,color,longtable}
\usepackage{array}
\newcolumntype{P}[1]{>{\centering\arraybackslash}p{#1}}
\usepackage{lscape}
\usepackage{multirow}
\usepackage{empheq}
\usepackage{comment}
\usepackage[utf8]{inputenc}
\usepackage{mathtools}
\usepackage{titlesec}
\usepackage{lipsum}
\usepackage{float}
\restylefloat{table}
\usepackage{physics}
\usepackage{xcolor}
\usepackage[skins,theorems]{tcolorbox}
\tcbset{highlight math style={enhanced,colframe=red,colback=yellow,arc=0pt,boxrule=1pt}}
\usepackage[makeroom]{cancel}

\newcommand{\tblue}{\textcolor{blue}}

\newlength{\extraspace}
\newlength{\extraspaces}
 \catcode`\@=11
\def\numberbysection{\@addtoreset{equation}{section}
\def\theequation{\arabic{section}.\arabic{equation}}}
\newcommand{\nonu}{\nonumber \\[.5mm]}

\newcommand{\be}{\begin{equation}}
\newcommand{\ee}{\end{equation}}
\newcommand{\bea}{\setlength\arraycolsep{2pt} \begin{eqnarray}}
\newcommand{\eea}{\end{eqnarray}}
\newcommand{\ba}{\begin{array}}
\newcommand{\ea}{\end{array}}
\newcommand{\bn}{\begin{align}}
\newcommand{\en}{\end{align}}

\newcommand{\eq}[1]{\eqref{#1}}
\newcommand{\nn}{\nonumber}
\newcommand{\w}[1]{\\[0.#1cm]}

\newcommand{\bpsi}{{\bar\psi}}
\newcommand{\eb}{{\bar\epsilon}}
\newcommand{\EA}{{\bar\epsilon}^A}

\usepackage[shadow,textwidth=2.7cm]{todonotes}
\usepackage{ifthen}
\reversemarginpar
\newcounter{todocounter}

\definecolor{aqua}{rgb}{0.0, 1.0, 1.0}
\definecolor{antiquewhite}{rgb}{0.98, 0.92, 0.84}
\definecolor{aquamarine}{rgb}{0.5, 1.0, 0.83}
\definecolor{bubblegum}{rgb}{0.99, 0.76, 0.8}
 	
\colorlet{escolor}{yellow!40!white}
\colorlet{ytcolor}{aquamarine!40!white}
\colorlet{hccolor}{bubblegum!40!white}

\newcommand{\esinline}[2][]{
  \ifthenelse { \equal {#1} {} }
    { \def\temp {#2} }  % if #1 == blank
    { \def\temp {#1} }   % else (not blank)
  \refstepcounter{todocounter}\todo[color=escolor,inline,caption={\textbf{\thetodocounter. ES} \temp}]{\textbf{\thetodocounter. ES:} #2}{}}

\newcommand{\ytinline}[2][]{
  \ifthenelse { \equal {#1} {} }
    { \def\temp {#2} }  % if #1 == blank
    { \def\temp {#1} }   % else (not blank)
\refstepcounter{todocounter}\todo[color=ytcolor,inline,caption={\textbf{\thetodocounter. YT} \temp}]{\textbf{\thetodocounter. YT:} #2}{}}

\newcommand{\hcinline}[2][]{
  \ifthenelse { \equal {#1} {} }
    { \def\temp {#2} }  % if #1 == blank
    { \def\temp {#1} }   % else (not blank)
\refstepcounter{todocounter}\todo[color=hccolor,inline,caption={\textbf{\thetodocounter. HC} \temp}]{\textbf{\thetodocounter. HC:} #2}{}}

\usepackage{multicol}
\newcommand{\al}{\alpha}
\newcommand{\del}{\delta}
\newcommand{\e}{\epsilon}
\newcommand{\g}{\gamma}

\newcommand{\la}{\lambda}

\newcommand{\m}{\mu}
\newcommand{\n}{\nu}

\newcommand{\pd}{\partial}
\newcommand{\rh}{\rho}
\newcommand{\s}{\sigma}
\newcommand{\ta}{\tau}

\newcommand{\vp}{\varphi}

\newcommand{\E}{{\mathcal E}}
\newcommand{\cL}{{\mathcal L}}
\newcommand{\cH}{{\mathcal H}}
\newcommand{\EW}{{  \widetilde{\cal E}}}

\makeatletter
\renewcommand{\theequation}{\thesection.\arabic{equation}}
\@addtoreset{equation}{section}
\makeatother

\allowdisplaybreaks
\begin{document}

{\flushright  MI-HET-799\\
STUPP-23-261 \\[15mm]}

\begin{center}
{ \Large \bf Higher derivative couplings of hypermultiplets }

\vspace{6mm}
\normalsize
{\large  Hao-Yuan Chang${}^{1}$, Ergin Sezgin${}^{1}$ and Yoshiaki Tanii${}^2$}

\vspace{10mm}

${}^4${\it Mitchell Institute for Fundamental Physics and Astronomy\\ Texas A\&M University
College Station, TX 77843, USA}

\vskip 1 em 

${}^2${\it Division of Material Science, Graduate School of Science and Engineering\\ Saitama University, Saitama 338-8570, Japan}

\vspace{10mm}

\hrule

\vspace{5mm}

\begin{tabular}{p{14cm}}

We construct the four-derivative supersymmetric extension of $(1,0), 6D$ supergravity coupled to Yang-Mills and hypermultiplets. The hypermultiplet scalars are taken to parametrize the quaternionic projective space $Hp(n)=Sp(n,1)/Sp(n)\times Sp(1)_R$. The hyperscalar kinetic term is not deformed, and the quaternionic K\"ahler structure and symmetries of $Hp(n)$ are preserved. The result is a three parameter Lagrangian supersymmetric up to first order in these parameters. Considering the case of $Hp(1)$ we compare our result with that obtained from the compactification of $10D$ heterotic supergravity on four-torus, consistently truncated to $N=(1,0)$, in which the hyperscalars parametrize $SO(4,1)/SO(4)$. We find that depending on how $Sp(1) \subset Sp(1,1)$ is embedded in  $SO(4)$, the results agree for a specific value of the parameter that governs the higher derivative hypermultiplet couplings.

\end{tabular}

\vspace{6mm}
\hrule
\end{center}

\newpage

\setcounter{tocdepth}{2}
\tableofcontents

\medskip

\section{Introduction}
%%%%%%%%%%%%%%%%%%%%%%%%%%%

There exists a class of matter coupled gauged supergravities in six dimensions which are anomaly free in a highly nontrivial fashion, and yet they do not seem to follow from string/M theory \cite{RandjbarDaemi:1985wc,Avramis:2005qt,Avramis:2005hc}. The anomaly cancellation in these theories requires Green-Schwarz mechanism in which the 3-form field strength needs to be modified to include a Lorentz Chern-Simons term which breaks supersymmetry. Restoring supersymmetry in this on-shell supergravity leads to a derivative expansion in its own right, without necessarily having a connection to string theory. As such, it is natural to study the higher derivative extensions of supergravity in the context of effective theory of quantum supergravity, as well as the swampland program, in which it would be useful to determine if such an effective theory passes the tests for providing consistent couplings of matter to supergravity. 

In this paper, we study the four-derivative extension of $N=(1,0)$ supergravity in six-dimensions coupled to Yang-Mills and hypermultiplets. We use the terminology of $N=(1,0)$ supergravity for short, to mean minimal supergravity coupled to a single tensor multiplet. 
Taking advantage of the fact that the $N=(1,0), 6D$ supergravity is known off-shell \cite{Bergshoeff:1985mz}, two independent curvature-squared off-shell invariants have been constructed \cite{Bergshoeff:1986vy,Bergshoeff:1986wc,Butter:2018wss}.

It has been shown that a specific combination, upon dualization gives the $6D$ analog of the Riemann-squared extension of heterotic supergravity in $10D$ which we refer to as Bergshoeff-de Roo (BdR) supergravity \cite{Chang:2022urm}  (see also \cite{Liu:2013dna}). Here, we shall consider this theory in $6D$, referring to it as BdR supergravity as well, since it has the same form as the corresponding supergravity in $10D$, together with Yang-Mills and hypermultiplet couplings. The YM coupling is straightforward, since the two derivative YM action arises at the same order as the Riemann-squared term, as in $10D$ supergravity as considered as low energy limit of heterotic string. The two-derivative coupling is also known. The challenge is to construct the four-derivative couplings in the hypermultiplet sector, and it seems that it has not been addressed adequately in the literature so far. Part of the difficulty stems from the fact that the hypermultiplets do not lend themselves to a simple off-shell description. In this paper we shall not rely on superspace and we will construct the higher-derivative hypermultiplet couplings to Einstein-Yang-Mills supergravity with Riemann-squared extension by employing the Noether procedure. 

As was first shown in \cite{Bagger:1983tt}, locally supersymmetric coupling of  hypermultiplets to supergravity requires that hyperscalars parametrize a quaternionic K\"ahler (QK) manifold of negative constant scalar curvature. We shall recall basic facts about such manifolds in the next section. In this paper we shall take the QK manifold to be the symmetric quaternionic projective space $Hp(n)= Sp(n,1)/ Sp(n)\times Sp(1)_R$, where $Sp(1)_R$ denotes the R-symmetry group. Denoting by $L$ the representative of this coset space, the Maurer-Cartan form $L^{-1} d L = P+Q_{Sp(n)} + Q_{Sp(1)_R}$, defines, as usual, the covariant derivative of the scalars, and the composite connections denoted by $Q$. In this paper, we shall consider the construction of a Lagrangian  of the form
\be
\cL= \cL(R)+ \cL(P^2) + \beta \cL(F^2) + \alpha \cL(R^2) +\cL_{\alpha,\gamma} (P^4)\ ,
\label{L}
\ee
where the first three terms denote the two derivative couplings of hypermultiplets and Yang-Mills to supergravity, $\cL(R^2)$ denotes the BdR action in $6D$ discussed above, and $\cL_{\alpha,\gamma} (P^4)$ denotes the four-derivative hypermultiplet couplings. The parameter $\beta\equiv 1/g^2_{YM}, \alpha$ and $\gamma$ are arbitrary parameters, which go like the inverse string tension $\alpha'$, if a string theory embedding of the model would exist. The main result of this paper is the determination of 
$\cL_{\alpha,\gamma} (P^4)$. We shall do so by Noether procedure, implementing supersymmetry up to first order in $\alpha, \beta, \gamma$. To that end, we parametrize the most general dimension four couplings of the hypermultiplet field to supergravity, requiring 33 parameters. As a result of Noether procedure we find that all of these parameters depend on $\alpha$ and $\gamma$, and supersymmetry is established up to first order in these parameters. The reason for presence of $\alpha$ dependent terms in $\cL_{\alpha,\gamma} (P^4)$ is due the  fact that the fermionic fields of supergravity necessarily couple to $Sp(n)\times Sp(1)_R$ connections, and as a result the supersymmetry variations of some of the $\alpha$ dependent terms conspire with some of the $\gamma$ dependent term to cancel. As a result, there will be some terms in the $\alpha$ dependent part of the Lagrangian that depend on the higher derivative hypermultiplet couplings (see, comments below \eq{FL}). 

The paper is organized as follows. In section 2, we recall the properties of quaternionic K\"ahler manifolds, and focus on the quaternionic projective spaces. In section 3, we describe the Noether procedure strategy we follow, and a general ansatz for the higher derivative hypermultiplet couplings. We also explain the construction of the Riemann-squared extension of $6D, N=(1,0)$ supergravity, which brings in an arbitrary parameter, $\alpha$. The Yang-Mills couplings are at the two-derivative level, and have the independent overall parameter $\beta= 1/g_{YM}^2$. In section 4, we carry out the Noether procedure, and determine all the parameters appearing in the ansatz for the total Lagrangian, and show that the hypermultiplet couplings bring in a single new constant, $\gamma$. In section 5, we compare the $Hp(1)$ truncation of our results with that of Riemann-squared extension of $10D$ heterotic supergravity on $T^4$, followed by suitable truncations. Our notations and conventions are given in appendix A, the lowest order field equations in appendix B, and several useful identities used in the Noether procedure in appendices C and D.

%%%%%%%%%%%%%%%%%%%%%%%%%%%%%%%%%%%%%%%%%%%%%%%%%%%%%%%%%%%%%%%%%%%%%%%%%%%%
\section{Hypermultiplets and quaternionic K\"ahler manifolds}
%%%%%%%%%%%%%%%%%%%%%%%%%%%%%%%%%%%%%%%%%%%%%%%%%%%%%%%%%%%%%%%%%%%%%%%%%%%%

\subsection{Generalities}
%%%%%%%%%%%%%%%%%%%%%%%%%%%%%%%%%%%%%%%%%%%%%%%%%

In 1983 Bagger and Witten \cite{Bagger:1983tt} showed that arbitrary number of hypermultiplets coupled to 
$N=2$ supergravity in $4D$ parametrise a quaternionic K\"ahler (QK) manifold with constant negative scalar curvature. A year 
later a similar result for the couplings of hypermultiplets to $N=(1,0)$ supergravity in $6D$ was presented in 
\cite{Nishino:1984gk}. The QK in question can be noncompact Wolf spaces, all of which are symmetric coset spaces, or 
 Alekseevsky spaces \cite{Alekseevsky'75,deWit:1991nm} which are homogeneous by non-symmetric cosets $G/H$ where $G$ is 
 not simple, or more  general QK manifolds which are not 
 homogeneous\footnote{We thank Guillaume Bossard for pointing these references to us.}  \cite{LeBrun:1991,Cortes:2020klb}.
The result for the full $N=(1,0), 6D$ supergravity coupled to arbitrary number of hypermultiplets parametrizing an arbitrary QK manifold 
can be found in \cite{Nishino:1984gk,Nishino:1986dc}, where the gauging of full isotropy group in the case of the noncompact Wolf space $Sp(n,1)/Sp(n)\times Sp(1)$,  was also given. 

In this section, we recall the result of \cite{Nishino:1984gk,Nishino:1986dc} for $N=(1,0), 6D$ supergravity, which has the field content
\be
\{ e_\m{}^r, B_{\m\n}, \vp\, ; \psi_\m^A, \chi^A \}\ ,
\ee
coupled to $n_H$  number of hypermultiplets with fields
\be
\{ \phi^\alpha, \psi^a \}\ ,\qquad a=1,...,2n, \qquad \alpha=1,...,4n\ .
\ee
The fermions $(\psi_\m^A, \chi^A, \psi^a)$ are symplectic-Majorana-Weyl, and $A=1,2$. The rest of the notation should be self-explanatory. Let us denote the vielbeins on the  QK manifold by $V_{\alpha}^{aA}$ and its inverse by $V^\alpha_{aA}$. They satisfy the relations \cite{Bagger:1983tt}
\be
g_{\alpha\beta} V^\alpha_{aA} V^\beta_{bB} = \Omega_{ab} \e_{AB}\ , \qquad V^{\alpha aA} V^\beta_{a B} + \alpha \leftrightarrow \beta = g^{\alpha\beta} \delta^A_B\ ,
\ee
where $g_{\alpha\beta}$ is the metric, and $\Omega_{ab}, \epsilon_{AB}$ are the antisymmetric  invariant tensors of $Sp(n)$ and $Sp(1)$, respectively. The vielbeins are covariantly constant,  and the triplet of complex structures $J^i, \ i=1,2,3$ obeying the quaternion algebra $[J^i,J^j]=\e^{ijk} J^k$ can be expressed as
\be
J_{\alpha\beta}^i =  T^i_A{}^B \Big( V_\alpha^{aA} V_{\beta aB} - \alpha \leftrightarrow \beta \Big) \ ,
\label{J}
\ee
where $T^i= -i\sigma^i/2$ are the $SU(2)$ generators. The integrability condition $[D_\alpha, D_\beta] V^\gamma_{aA}=0$ gives \cite{Bagger:1983tt}
\be
R_{\alpha\beta\gamma\delta} V^\delta_{aA} V^\gamma_{bB} = \e_{AB} Q_{\alpha\beta ab} +\Omega_{ab} Q_{\alpha\beta AB}\ ,
\ee
where $Q_{\alpha\beta ab}$ and $Q_{\alpha\beta AB}$ are the $Sp(n_H)$ and $Sp(1)$ valued curvatures, respectively. The cyclic identity for $R_{\alpha\beta\gamma\delta}$ and \eq{J} imply that
\be
Q_{\alpha\beta ab} = \kappa^2 \Big( V_{\alpha aA} V_{\beta b}{}^A - \alpha \leftrightarrow \beta \Big) + \Omega_{abcd} V_\alpha^{dA} V_\beta{}^c{}_A\ ,
\ee
where $\Omega_{abcd}$ is a totally symmetric $Sp(n)$ tensor, and $Q_{\alpha\beta A}{}^B= -2 V_{[\alpha}{}^{Ba} V_{\beta] a A}$.  

The complete action that describes the coupling of an arbitrary QK sigma model to $6D, N=(1,0)$ supergravity was constructed in \cite{Nishino:1986dc}\footnote{Only in the context of gauging isometries of the QK space that the quaternionic projective space $G/H= Sp(n,1)/[S(n)\times Sp(1)]$ was picked in particular, and the group $H$ was gauged.}
in a fashion similar to that of Bagger and Witten \cite{Bagger:1983tt}. The geometrical ingredients  described above are key to this construction, even though it should be noted that the $Sp(n)$ tensor $\Omega_{abcd}$ arise only in the quartic fermion term
\begin{align}
-\tfrac{1}{18} \Omega_{abcd}\, \bpsi^a \gamma_\m \psi^b \bpsi^c \gamma^\m \psi^d\ .
\end{align}

\subsection{The case of $Hp(n)$} 
%%%%%%%%%%%%%%%%%%%%%%%%%%%%%%%%%%%%%

In this section we shall take the QK manifold parametrized by the hyperscalars to be the 
quaternionic projective space $Hp(n)$ with $n>1$,  which can be realized as the coset  $Sp(n,1)/[Sp(n)\times Sp(1)]$,  that has real dimension $4n$. In this case the tensor 
\be
\Omega_{abcd}=0\ .
\ee
Using the $(2n+2) \times (2n+2)$ matrix $L$ of $Sp(n,1)$ as a representative of the coset, 
the Maurer-Cartan form can be written as
\begin{equation}
L^{-1} dL = \left( 
\begin{array}{cc}
Q_a{}^b & P_a{}^B \\
P_A{}^b & Q_A{}^B
\end{array}
\right), 
\label{oneform}
\end{equation}
where $ Q_{ab} = Q_{ba},  Q_{AB} = Q_{BA},\ P_{Ab} = -P_{bA}$ and 
\be
P_\m^{aA} = \partial_\m \phi^\alpha\,V_\alpha^{aA}\ ,\qquad Q_\m^{AB} 
= \partial_\m \phi^\alpha\, Q_\alpha^{AB}\ ,\qquad 
Q_\m^{ab} = \partial_\m \phi^\alpha\, Q_\alpha^{ab}\ .
\ee
Note that $X := L^{-1}dL$ is a general element of the Lie algebra $Sp(n,1)$, and therefore it satisfies the condition $(\Omega X)^T= \Omega X$ where 
\begin{equation}
\Omega = \left( 
\begin{array}{cc}
\epsilon^{ab} & 0 \\
0 & - \epsilon^{AB}
\end{array}
\right). 
\end{equation}
The Maurer-Cartan equation $ d(L^{-1}dL) + L^{-1}dL \wedge L^{-1}dL = 0$ gives 
\begin{align}
Q_{\mu\nu a}{}^b 
&:= 2 \partial_{[\mu} Q_{\nu]a}{}^b + 2 Q_{[\mu|a}{}^c Q_{|\nu]c}{}^b 
= 2 P_{[\mu|a}{}^C P_{|\nu]}{}^b{}_C\ , 
\nn\w2 
Q_{\mu\nu A}{}^B 
&:= 2 \partial_{[\mu} Q_{\nu]A}{}^B + 2 Q_{[\mu|A}{}^C Q_{|\nu]C}{}^B 
= 2 P_{[\mu|}{}^c{}_A P_{|\nu]c}{}^B\ , 
\nn\w2 
D_{[\mu} P_{\nu]a}{}^B 
&:= \partial_{[\mu} P_{\nu]a}{}^B 
+ Q_{[\mu|a}{}^c P_{\nu]c}{}^B + Q_{[\mu}{}^{BC} P_{\nu]aC} = 0\ .
\label{mc}
\end{align}
With this building blocks, the locally supersymmetric two derivative  QK sigma model \cite{Nishino:1984gk,Nishino:1986dc} can be adapted to $Hp(n)$, and with field redefinitions \eq{sf} can be applied, to pass to the string frame. 

\section{The Noether procedure}
%%%%%%%%%%%%%%%%%%%%%%%%%%%%%%%%%

\subsection{Strategy}
%%%%%%%%%%%%%%%%%%%%%%%%%%%%%%%%%%%%%%%%%%%%%%%%%%%%%

We adopt the following strategy for the Noether procedure. For the purposes of the present discussion, it is convenient to write the total Lagrangian as $\cL= \cL_0 + \cL_1$ where $\cL_0$ represents the two-derivative\footnote{In referring to $n$-derivative couplings, we mean the bosonic sector, while it is $(n-1)$-derivative coupling in the fermionic sector.} part of supergravity coupled to hypermultiplets, and $\cL_1$ is the four-derivative extension plus the two-derivative couplings of the Yang-Mills multiplet $(A_\m^I, \lambda^I)$. Putting a generic small parameter in front of $\cL_1$, the supersymmetry variation of the action up to first order in that parameter takes the form
\begin{align}
\delta I &= \int d^6 x\, \Big( \delta_0 \cL_0 + \delta_0 \cL_1 + \delta_1 \cL_0 \Big)
\nn\w2
&= \int d^6 x\, \Big( \delta_0\cL_1 + \frac{\delta\cL_0}{\delta\phi} \delta_1\phi \Big) = \int d^6 x\, \Big( \delta_0\cL_1 + \E_\phi \delta_1\phi \Big)\ ,
\end{align}
where $\phi$ schematically denotes the set of fields in the theory, $\E_\phi$ denotes their field equations that follow from the lowest order action, and we have used the fact that $ \int d^6 x\, \delta_0 \cL_0=0$. Thus the invariance of the action at first order requires that $\int d^6 x\, \delta_0 \cL_1$ vanishes up to lowest order field equations, to wit
\be
\int d^6 x\, \delta_0 \cL_1 = \int d^6 x\, f(\e,\phi) \E_\phi\ ,
\label{dL1}
\ee
where $f(\e,\phi)$ is a functional of the fields, possibly containing $\E_\phi$ factors, and the supersymmetry parameter, possibly including its derivative. It then follows that supersymmetry is ensured by letting
\be
\delta_1\phi = -f(\e,\phi)\ .
\ee
In the Noether procedure, consideration of the $H=dB$ and $\vp_\m \equiv \partial_\m \vp$ independent variations of the action to begin with is motivated by the expectation that supersymmetry is powerful enough to establish $\cL_1$ even by consideration of such variations alone. In constructing $\cL_1$, we shall parametrize the most general four-derivative terms that include hypermultiplet fields, such that we omit terms proportional to EOM's that follow from $\cL_0$, since they automatically satisfy \eq{dL1}. Once the vanishing of $H$ and $\vp_\m$ independent variations are established, we then turn to the $H$ and $\vp_\mu$ dependent variations as well, and determine if new terms need to be added to the Lagrangian.

\subsection{The ansatz}
%%%%%%%%%%%%%%%%%%%%%%%%%%%%

An action with the two derivative couplings of hypermultiplets and Yang-Mills multiplets \cite{Nishino:1986dc} can be extended readily by introducing the ${\rm Riem}^2$ terms similar to the one in 10D \cite{Bergshoeff:1989de}. Let us denote the Lagrangian for this system as
\be
\cL= \underbrace{\cL(R)+ \cL(P^2)}_{\cL_0} + \beta \cL(F^2) + \alpha \cL(R^2)\ ,
\ee
where $\cL(R)$ is the $(1,0)$ supergravity Lagrangian, $\cL(P^2)$ is the Lagrangian that describes the two-derivative couplings of hypermultiplets, $\cL(F^2)$ describes the couplings of Yang-Mills multiplets, and $\cL(R^2)$ is the Bergshoeff-de Roo type higher derivative extension, derivation of which will be given at the end of this section.  In the spirit of heterotic supergravity, we will treat the constant parameters $\alpha$ and $\beta$ to be at the same footing in an expansion scheme in  these parameters. The Lagrangians $\cL_0, \cL(F^2)$ and $\cL(R^2)$ are given below in \eq{LP2}, \eq{F22} and \eq{R2}, respectively. 

Our goal is to extend this Lagrangian to describe four derivative couplings of the hypermultiplets. Thus we consider a Lagrangian of the form
\be
\cL= \cL_0 + \beta \cL(F^2) + \alpha \cL(R^2) +\cL_{\alpha,\gamma} (P^4)\ ,
\label{Lag}
\ee
where  $\cL(P^4)$ represents the higher derivative couplings of the hypermultiplets. The fact that the higher derivative hypermultiplet couplings turn out to depend only on $\gamma$ and $\alpha$ is a nontrivial consequence of the Noether procedure.  In $\cL(P^4)$ we have allowed dependence on not just a new coupling constant $\gamma$ but also dependence on $\alpha$ because $\alpha \cL(R^2)$ has gravitino curvature terms in which the  covariant derivative contains the composite connection which is a function of the hyperscalars. The hypermultiplet dependent terms in the variation of $\cL(R^2)$, given below in \eq{R2}, add up to
\begin{align}
\delta_0 \cL(R^2)\Big|_{\rm hypers}  &= e e^{2\vp} \Big[ -2( \EA \slashed{D} \psi_{\m\n}^B ) \, Q^{\m\n}{}_{AB}
+ \Big( \tfrac12  \EA \g^{\m\n\tau} \psi_\tau^B 
-\EA \g^\m \psi^{\n B} + \EA \g^{\m\n} \chi^B \Big)R_{\m\n}{}^{\rh\s} Q_{\rh\s AB}\Big] \ . 
\label{3t}
\end{align}
Using \eq{id1} in this expression, which is a consequence of the gravitino field equation, gives
\begin{align}
\delta_0 \cL(R^2) &= e e^{2\vp} \Big[-( \eb \g^\m \psi_\m ) Q^2
-2 ( \eb \chi ) Q^2  -4( \EA \g_\m \psi_\n^B ) \big( Q^{\m\rh} Q^\n{}_\rh \big)_{(AB)} 
\nn\w2
& +2( \eb \g^\m \psi^\n ) \big( Q_{\m\rh} Q_\n{}^\rh \big) 
+4 ( \EA \g_\m \psi_\n^B ) (P^2)^{\m\rh} Q^\n{}_{\rh AB} +8 ( \EA D_\m \psi^a ) P_{\n a}{}^B Q^{\m\n}{}_{AB}\Big]\ ,
\end{align}
where $Q^2 := Q_{\m\n AB} Q^{\m\n AB}$, and we have set to zero the equations of motion $\E_{\m\n}, \E_\m^A$ and $\E^A$, discussed in appendix B.  These terms trigger the Noether procedure which requires the addition of higher derivative hypermultiplet dependent terms. We have opted for adding the most general such terms as detailed in the section below.

The first three terms in \eq{Lag} are known and, up to quartic fermion terms, they are given by  (see appendix A for the definitions of notations)
\begin{align}
\cL_0
&= e e^{2\varphi} \biggl[ \, \tfrac{1}{4} R(\omega) + \partial_\m\vp \partial^\m \vp
- \tfrac{1}{12} H_{\mu\nu\rho} H^{\mu\nu\rho} -\tfrac12 P_\mu^{aA} P^\mu_{aA}
\nn\\
& - \tfrac{1}{2} \bar{\psi}_\mu \gamma^{\mu\nu\rho} D_\nu \psi_\rho 
+ 2 \bar{\chi} \gamma^{\mu\nu} D_\mu \psi_\nu 
+ 2 \bar{\chi} \gamma^\mu D_\mu \chi - \tfrac{1}{2} \bar{\psi}^a \gamma^\mu D_\mu \psi_a
\nn\w2
&    - \tfrac{1}{24} H_{\mu\nu\rho} {\cal O}^{\m\n\rho} 
- \partial_\mu \vp \left( 
\bar{\psi}^\mu \gamma^\nu \psi_\nu 
+ 2 \bar{\psi}_\nu \gamma^\mu \gamma^\nu \chi \right) 
 -P_{\mu aA} \left( \bar{\psi}_\nu^A \gamma^\mu \gamma^\nu \psi^a + 2 \bar{\chi}^A \gamma^\mu \psi^a \right)\biggr]\ ,
\label{LP2}\w2
\cL(F^2)
&= e e^{2\varphi} \biggl[ \, 
- \tfrac{1}{4} F_{\mu\nu}^I F^{I\mu\nu}
- \bar{\lambda}^I \gamma^\mu D_\mu \lambda^I - \tfrac{1}{12} H_{\mu\nu\rho} 
\bar{\lambda}^I \gamma^{\mu\nu\rho} \lambda^I
+ \tfrac{1}{2} F_{\mu\nu}^I \bar{\lambda}^I \left( 
\gamma^\rho \gamma^{\mu\nu} \psi_\rho + 2 \gamma^{\mu\nu} \chi \right) 
\nn\w2
& \qquad\qquad  + \omega^{YM}_{\m\n\rh} \big( H^{\m\n\rh} +\tfrac14 {\cal O}^{\m\n\rh} \big)        \biggr]\ ,
\label{F22}\w2
\cL(R^2)
&= e e^{2\varphi} \biggl[ \, 
- \tfrac{1}{4} R_{\mu\nu}{}^{rs}(\Omega_-) R^{\mu\nu}{}_{rs}(\Omega_-) 
- \bar{\psi}^{rs} \gamma^\mu D_\mu (\omega,\Omega_-)\,  \psi_{rs}
\nn\\
& \qquad - \tfrac{1}{12} H_{\mu\nu\rho} 
\bar{\psi}^{rs} \gamma^{\mu\nu\rho} \psi_{rs} + \tfrac{1}{2} R_{\mu\nu}{}^{rs}(\Omega_-) \, \bar{\psi}_{rs} 
\left( \gamma^\rho \gamma^{\mu\nu} \psi_\rho + 2 \gamma^{\mu\nu} \chi 
\right) 
\nn\w2
& \qquad\qquad  + \omega^{L}_{\m\n\rh} \big( H^{\m\n\rh} +\tfrac14 {\cal O}^{\m\n\rh} \big)   \biggr] \ ,
\label{R2}\w2
\cL_{\alpha,\gamma} (P^4)
&= e\,e^{2\varphi} \Bigg[ \Big( b_1\, Q^2 + b_2 (P^2)_{\mu\nu} (P^2)^{\mu\nu} + b_3 (P^2)^2  \Big)
\nn\w2
& + \Big(\, c_1\,\bar{\psi}_\rh^A \gamma^\rh \psi^{\m\n B} +c_2\,\bar{\chi}^A \psi^{\m\n B} + c_3\, \bpsi_\rh^A \gamma^\m \psi^{\rh\n B}\Big) Q_{\m\n AB} 
+ c_4\,\bpsi_\rh \gamma^\m \psi^{\n\rh} (P^2)_{\mu\nu} 
\nn\w2
& + c_5 \bpsi^a \gamma_\mu D_\nu \psi^b (P^2)^{\mu\nu}_{ab} +  c_6\, \bpsi \gamma_\mu D_\nu \psi (P^2)^{\mu\nu} 
\nn\w2
& +  \Big( c_7\,\bpsi_\mu^A \gamma^{\mu\nu\rho} \psi_\rho^B
+ c_8\,\bpsi_\mu^A \gamma^{\mu\nu} \chi^B  +c_9\,\bpsi^{\nu A} \chi^B
 \Big) (PDP)_{\nu AB} 
\nn\w2
& + \Big( c_{10}\,\bpsi^\nu \gamma^\mu \psi^\rho+c_{11}\,\bpsi^\mu \gamma^{\nu\rho} \chi \Big) (PDP)_{\mu,\nu\rho} + c_{12}\, \bpsi^a \gamma_\mu \psi^b (PDP)^\mu_{ab} 
\nn\w2
&  
+c_{13}\,\bpsi_\rho^A\gamma_{\m\n}\chi^B\, D^\rho Q^{\m\n}{}_{AB}
+\Big(\,c_{14}\, \bpsi^\m \gamma^\n \psi_\n +c_{15}\,\bpsi^\mu \chi \Big)\, \partial_\m P^2 
\nn\w2
& +  {\bar\chi}_A \gamma^\m \psi_a \Big( c_{16}\,Q_{\m\n}^{AB} P^{\n a}{}_B   + c_{17}\, (P^2)_{\m\n} P^{\n aA}   +  c_{18}\, P_\mu^{aA} P^2 \Big)
\nn\w2
&  +\bpsi_\mu^A \psi_a  \Big(c_{19}\, Q^{\mu\nu}_{AB} P_\nu^{aB} + c_{20}\, (P^2)^{\mu\nu} P_\nu^a{}_A 
+c_{21} P^{\mu a}{}_A P^2 \Big)  
\nn\w2
& + \bpsi_{\mu A} \gamma^{\mu\nu} \psi^a \Big( c_{22}\, Q_{\nu\rho}{}^{AB} P^\rho{}_{aB} + c_{23}\, (P^2)_{\nu\rho} P^\rho{}_a{}^A +c_{24} P_{\n a}{}^A P^2 \Big)  
\nn\w2
& +  \bpsi_\mu^A \gamma_{\nu\rho} \psi_a \Big( c_{25}\, Q^{\mu\nu}_{AB} P^{\rho aB} +  c_{26}\, Q^{\nu\rho}_{AB} P^{\mu aB} + c_{27}\, (P^2)^{\mu\nu} P^{\rho a}_A  +c_{28}\, R^{\mu\nu\rho\sigma} P_\sigma{}^a{}_A \Big) 
\nn\w2
&  + c_{29}\, {\bar\chi}^A \gamma^{\mu\nu\rho} \psi^a  Q_{\m\n\, A}{}^B P_{\rh aB} 
+ c_{30}\,{\bar\psi}_\mu^A \gamma^{\mu\nu\rho\sigma} \psi^a\, Q_{\nu\rho A}{}^B P_{\sigma aB} 
\nn\w2
& + \gamma \omega^Q_{\m\n\rh} \big( H^{\m\n\rh} +\tfrac14 {\cal O}^{\m\n\rh} \big) \biggr]\ ,
\label{L3c}
\end{align}
where
\allowdisplaybreaks{
\begin{align}
H_{\m\n\rho} &= 3\partial_{[\m} B_{\n\rh]}\ ,
\nn\w2
\Omega_{\pm \mu rs} &= \hat{\omega}_{\mu rs} \pm \hat{H}_{\mu rs}\ , 
\nn\w2
\hat{\omega}_{\mu rs} &= \omega_{\mu rs} 
+ \tfrac12 \left( \bar{\psi}_\mu \gamma_r \psi_s 
- \bar{\psi}_\mu \gamma_s \psi_r
+ \bar{\psi}_r\gamma_\mu \psi_s \right)\ ,
\nn\w2
\hat{H}_{\mu\nu\rho} &= H_{\mu\nu\rho} 
+ \tfrac32 \bar{\psi}_{[\mu} \gamma_\nu \psi_{\rho]}\ , 
\nn\w2
\psi_{\m\n}^A &= \Big(\left( \partial_\m  + \tfrac{1}{4} \Omega_{+\m rs} \gamma^{rs}\right) \psi_\n^A +Q_\m{}^{AB} \psi_{\n B}\Big) - \m\leftrightarrow \n\ .
\label{def1}
\end{align}
Furthermore we have the Chern-Simon forms
\begin{align}
\omega^{YM}_{\m\n\rh} &=  \tr \Big( A_{[\m}\partial_\n A_{\rh]} + \tfrac23 A_{[\m} A_\n A_{\rh]}\Big)\ ,
\nn\w2
\omega^L_{\m\n\rh} &=  \tr \Big( \Omega_{-[\m}\partial_\n \Omega_{-\rh]} + \tfrac23 \Omega_{-[\m} \Omega_{-\n} \Omega_{-\rh]}\Big)\ ,
\nn\w2
\omega^Q_{\m\n\rh} &=  \tr \Big( Q_{[\m}\partial_\n Q_{\rh]} + \tfrac23 Q_{[\m} Q_\n Q_{\rh]}\Big)\ ,
\nn\w2
 &= \Big( Q_{\m A}{}^B \partial_\n Q_{\rho B}{}^A +\tfrac23 Q_{\m A}{}^B Q_{\nu B}{}^C Q_{\rho C}{}^A \Big)_{[\m\n\rho]}\ ,
 \label{CSQ}
\end{align}
where $A_\m := A_\m^I T^I$ and $\tr (T^I T^J) = -\delta^{IJ}$. We have anticipated that the Chern-Simons term for the composite connection on $Hp(n)$ will be needed in the Noether procedure. At this point there is no loss of generality in doing so since we have introduced an arbitrary coupling constant $\gamma$ in front of it.  Further definitions are the fermionic bilinear terms
\begin{align}
{\cal O}_{\m\n\rh} &= 
\bar{\psi}^\sigma \gamma_{[\sigma} \gamma^{\mu\nu\rho} 
\gamma_{\tau]} \psi^\tau 
+ 4 \bar{\psi}_\sigma \gamma^{\sigma\mu\nu\rho} \chi 
- 4 \bar{\chi} \gamma^{\mu\nu\rho} \chi + \bar{\psi}^a \gamma^{\mu\nu\rho} \psi_a\ , 
\end{align}
and the covariant derivatives which now contain the $Sp(n)\times Sp(1)$ connections,
\begin{align}
D_\m \chi^A &= \big(\partial_\m + \tfrac14 \omega_\m{}^{rs} \gamma_{rs} \big)\chi^A + Q_\m{}^{AB} \chi_B\ ,
\nn\\
D_\m \psi^a &= \big(\partial_\m + \tfrac14 \omega_\m{}^{rs} \gamma_{rs} \big)\psi^a + Q_\m{}^{ab} \psi_b\ .
\end{align}
The coefficients $b_1, b_2, b_3, c_1,..., c_{30}$ will turn out to be linear in $\gamma$ and $\alpha$. Note also that the Chern-Simons terms in $\cL(F^2), \cL(R^2), \cL_{\alpha,\gamma}$ can be absorbed into the definition of $H=dB$ to define $\cH$ as follows
\be
\cH_{\m\n\rh} = 3 \partial_{[\mu} B_{\nu\rh]} - 6 \beta \omega^{YM}_{\m\n\rh} - 6 \alpha\,\omega^L_{\m\n\rh}
- 6 \gamma \,\omega^Q_{\m\n\rh}\ .
\label{CH}
\ee

In the ansatz for $\cL_{\al,\g}(P^2)$, we have assumed that the derivative of the gravitino appears only through the gravitino curvature. This is motivated by dimensional reduction of the $R+\alpha\, {\rm Riem}^2$ action in $10D$ on $T^4$ that was carried out in \cite{Chang:2021tsj}. This reduction also gives a term of the form $\left(D^\m P_\n^{aA} \right)^2$. However, in this case we have opted to parametrize the four-derivative hyperscalar terms as in \eq{L3c}, in view of the following identity
\begin{align}
\int d^6 x\, ee^{2\vp} \left(D^\m P_\n^{aA} \right) \left(D_\m P^\n_{aA} \right) =&\, \int d^6 x\, ee^{2\vp} \big[ -2 (P^2)_{\m\n} (P^2)^{\m\n} -\tfrac12 Q_{\m\n ab} Q^{\m\n ab} 
\nn\w2
& -\tfrac12 Q_{\m\n AB} Q^{\m\n AB} - P^{\m a A} D_\m D^\n P_{\n a A} - (P^2)^{\m\n} R_{\m\n} \big]\ .
\end{align}
Removing the last three terms by using the field equations, and using \eq{C5} as well, leads to the ansatz \eq{L3c} with redefined parameters. 

Turning to the supersymmetry transformation rules, in accordance with the Noether procedure strategy outlined above, we need to start with the following ones: 
\begin{align}
\delta e_\mu{}^m &= \bar{\epsilon} \gamma^m \psi_\mu\ , 
\nn\\
\delta \psi_\mu &= D_\mu \epsilon 
+ \tfrac{1}{4} \cH_{\mu\rho\sigma} \gamma^{\rho\sigma} \epsilon\ , 
\nn\\
\delta B_{\mu\nu} 
&= - \bar{\epsilon} \gamma_{[\mu} \psi_{\nu]} 
+2 \beta\, A_{[\m}^I \delta A_{\n]}^I + 2 \alpha\,\Omega_{-[\mu}{}^{rs}\delta\Omega_{-\nu] rs}
+ 2 \gamma\, Q_{[\mu}{}^{AB}\delta Q_{\nu] AB}\ , 
\nn\\
\delta \chi 
&= \tfrac{1}{2} \gamma^\mu \epsilon \partial_\mu  \varphi 
- \tfrac{1}{12} \cH_{\mu\nu\rho} \gamma^{\mu\nu\rho} \epsilon\ ,
\nn\\
\delta \varphi &= \bar{\epsilon} \chi\ ,
\nn\\
L^{-1} \delta L &= \left( 
\begin{array}{cc}
0 & - \bar{\epsilon}^B \psi_a \\
+ \bar{\epsilon}_A \psi^b & 0
\end{array}
\right) ,
\nn\\
\delta \psi^a &= - P_\m^{aA} \gamma^\m \e_A\ ,
\nn\\
\delta A_\mu^I &= - \bar{\epsilon} \gamma_\mu \lambda^I, \nonu 
\delta \lambda^I
&= \tfrac{1}{4} F_{\mu\nu}^I \gamma^{\mu\nu} \epsilon\ .
\label{super2}
\end{align}
Substituting the expression for $L^{-1}\delta L$ into the formula 
\begin{equation}
\delta ( L^{-1} dL ) 
= d(L^{-1}\delta L) + [ L^{-1}dL, \, L^{-1} \delta L]\ ,
\end{equation}
we find 
\begin{align}
\delta Q_\mu{}^{AB} 
&= 2\bar{\epsilon}^{(A|} \psi_c P_\mu{}^{c|B)}\ ,
\nn\w2
\delta Q_\mu{}^{ab} 
&= 2 \bar{\epsilon}_A \psi^{(a} P_\mu{}^{b) A}\ ,
\nn\w2
\delta P_\mu{}^{aA} 
&= - D_\mu \big(\bar{\epsilon}^A \psi^a \big) \ .
\label{dpq}
\end{align}
The gravitino curvature transforms under supersymmetry as 
\begin{align}
\delta \psi_{rs}^A
&= \tfrac{1}{4} R_{\mu\nu rs}(\Omega_-) \gamma^{\mu\nu} \epsilon^A
+Q_{rs}{}^{ AB} \e_B\ ,
\label{deltapsi}
\end{align}
which contains hypermultiplet dependent terms. The requirement of cancelling this variations triggers  the Noether procedure for constructing the four-derivative couplings of the hypermultiplets to supergravity. 

We end this section with some comments on the Lagrangians $\cL_0$ and $\cL(R^2)$. The Lagrangian $\cL_0$ was given completely in \cite{Nishino:1986dc} in Einstein frame. Here we have passed to the `string' frame by performing the field redefinitions
\begin{align}
e_\mu{}^r\ &\longrightarrow\ 
e^{\frac{1}{2}\varphi} e_\mu{}^r\ ,  \qquad     \psi_\mu\  \longrightarrow\  e^{\frac{1}{4}\varphi} \left( \psi_\mu 
+ \frac{1}{2} \gamma_\mu \chi \right), \nonu
\chi\ &\longrightarrow\ e^{-\frac{1}{4}\varphi} \chi\ , \qquad\ \    \psi^a\ \longrightarrow\ 
e^{-\frac{1}{4}\varphi} \psi^a\ ,\qquad  \epsilon\ \longrightarrow\ e^{\frac{1}{4}\varphi} \epsilon\ , \nonu
\delta(\epsilon) + \delta_L(\lambda) &
\longrightarrow \delta(\epsilon), \qquad \qquad
\lambda^m{}_n = \frac{1}{2}\bar{\epsilon}\gamma^m{}_n \chi  \ . 
\label{sf}
\end{align}
Note in particular the shift in the gravitino, and the Lorentz transformations with the field dependent parameter given in the last equation. The latter is needed to put in to a canonical form the supersymmetry transformation of the vielbein. We have also different conventions here, which are related to those of \cite{Nishino:1986dc}, as described in appendix A.

The Lagrangian $\cL(R^2)$ has already been discussed in \cite{Chang:2022urm, Chang:2021tsj}, in the absence of hypermultiplets. Here, we shall explain its derivation which is based on the observation that the fields $(\Omega_{-\mu}{}^{rs},\psi^{rs})$ transform under supersymmetry (to lowest order in $\alpha$) in fashion similar to the Yang-Mills multiplet fields $(A_\m^I, \lambda^I)$. More precisely, one finds that under supersymmetry\footnote{In obtaining $\delta\psi_{rs}$ , one uses the  identity  \cite{Bergshoeff:1986wc,Bergshoeff:1989de}
$ R_{pqrs}(\Omega_+) - R_{rspq}(\Omega_-)  = 4 D_{[p}(\omega) H_{qrs]}$. }
\bea
\delta \Omega_{-\mu rs} 
&=& - \bar{\epsilon} \gamma_\mu \psi_{rs} \ , 
\nn\w2
\delta \psi_{rs} &=& \frac{1}{4} R_{\mu\nu rs}(\Omega_-) \gamma^{\mu\nu} \epsilon \ ,
\label{yma}
\eea
which shows that ($\Omega_{-\mu rs}$, $\psi_{rs}$)  transform as the Yang-Mills multiplet fields $(A_\mu^I,\  \lambda^I ) $  valued in the fundamental representation of the Lorentz algebra. The well known coupling of Yang-Mills multiplet to supergravity then makes it possible to immediately write down the 
supersymmetrization of the $R+ \alpha R_{\m\n\rh\s}R^{\m\n\rh\s}$ up to order $
\alpha$ by employing the map
\be
\mbox{YM to Lorentz map:} \qquad  (A_\mu^I,\  \lambda^I ) \quad \to \quad (\Omega_{-\mu}{}^{rs}, \ \psi^{rs})\ .
\label{map}
\ee
This map applied to the well-known Yang-Mills coupled to supergravity, immediately yields $\cL(R^2)$ given in \eq{R2}. Note the somewhat unusual covariant derivative in which the connection $\Omega_-$ only acts on the vector index of the gravitino curvature as follows
\be
D_\mu(\omega, \Omega_-) \psi_{rs} = \left( \partial_\mu 
+ \tfrac{1}{4} \omega_{\mu pq} \gamma^{pq} \right) \psi_{rs} 
+ \Omega_{-\m r}{}^p \psi_{ps} + \Omega_{-\m s}{}^p \psi_{rp}\ .
\label{gc1}
\ee
Despite the fact that the coupling of Yang-Mills multiplet is exactly supersymmetric, the map \eq{map} provides an action invariant only up to order $\alpha$ because unlike $A_\m^I$  the field $\Omega_\mu{}^{rs}$ is not an independent field, but rather a function of the vielbein  and the $H$-field. Note also that introduction of the hypermultiplets requires the introduction of the $Sp(n)\times Sp(1)_R$ composite connections in the covariant derivatives of $(\psi_\m, \chi)$. Consequently, $\cL(R^2)$ as given in \eq{R2} is no longer a supersymmetric extension of $\cL(R)+\cL(P^2)$. By introducing appropriate $\alpha$ dependent terms in $\cL_{\alpha,\gamma}$ the supersymmetry will be restored.

\section{Supersymmetry variations of the total action}
%%%%%%%%%%%%%%%%%%%%%%%%%%%%%%%%%%%%%%%%%%%%%%%%%%%%%%%%

\subsection{Variations independent of $\vp_\mu$ and $H$}
%%%%%%%%%%%%%%%%%%%%%%%%%%%%%%%%%%%%%%%%%%%%%%%%%%%%%%%%%%%%%%%

There is no unique way of choosing a basis for the independent structures that need to vanish for supersymmetry. In what follows,  the coefficients collected in front of the chosen basis, and they all need to vanish. Of those, 16 of them contain the hyperino, 14 of them contain the dilatino, and  28 of them contain the gravitino. All of these structures are displayed below. They must vanish, and therefore the supersymmetry of the 35 parameter Lagrangian gives 58 equations for the 35 parameters, which is a highly nontrivial over-constrained system to admit a solution. A very long calculation yields the following results. 

% We have set $d_0=d_1=1$ and $d_2=1/2$ below. 

Collecting the independent structures, gives for the supersymmetry variations of \eq{Lag} which contain the hyperino the result
\begin{align}
V(\psi)
&= \tblue{( {\bar\e}^A \psi^a )} \Big[  \Big( 16 b_1  + \tfrac12 c_5 - 2 c_7 - c_{12} + 2 c_{19} \Big)  P_{\mu a}{}^B (PDP)^\mu_{AB} 
\nn\\
& + \Big( 2 b_2 + 4 b_3  + \tfrac18 c_5 + \tfrac12 c_6 - \tfrac14 c_{12} + \tfrac12 c_{20} + c_{21} - 2 c_{14} \Big) P_{\mu a A} \partial^\mu P^2 
\nn\w2
& +\Big( - c_4 + 4 b_2  + \tfrac14 c_5 + \tfrac12 c_{12} + c_{20} \Big)  (P^2)_{\mu\nu} D^\mu P^\nu_{aA} \Big]
\nn\\
& +\tblue{( {\bar\e}^A \gamma_{\mu\nu}\psi^a )}\Big[ \Big(  \tfrac18 c_5 - \tfrac14 c_{12} + c_{26} + \tfrac12 c_{22} \Big) P^\lambda{}_a{}^B D_\lambda Q^{\mu\nu}{}_{AB}
\nn\\
& +\Big( \tfrac14 c_5 + c_3 + \tfrac12 c_{12} - c_{25} + c_{22} \Big) Q^{\mu\lambda}{}_{AB} D^\nu P_{\lambda a}{}^B 
\nn\\
& +\Big( \tfrac12 c_6 - \tfrac12 c_{27} - c_{24} \Big)\,P^\m_{a A} \pd^\n P^2
\nn\\
& + \Big( - c_4 + \tfrac14 c_5 + \tfrac12 c_{12} + c_{27} - c_{23} \Big) (P^2)^{\mu\lambda} D^\nu P_{\lambda aA} 
\nn\\
& +\Big( \tfrac14 c_5 - \tfrac12 c_{12} - c_{23} + 2 c_{28} \Big) P_{\lambda aA} (PDP)^{\lambda,\mu\nu}  
\nn\\
& + \Big( - 2 c_8 - 2 c_{25} \Big)\, P^\mu{}_a{}^B  (PDP)^\nu_{AB} 
\Big] 
\nn\\
& +\tblue{( \EA D_\mu \psi^a )} \Big[ \Big( 8 \al + 4 c_1 - \tfrac12 c_5 + c_3 + c_{19} - c_{22} \,\Big) P_{\nu a}{}^B Q^{\mu\nu}_{AB}
\nn\\
& +\Big( \tfrac12 c_5 + c_{21} + c_{24} \Big) P^\mu_{aA} P^2
+\Big( c_4 + \tfrac12 c_5 + 2 c_6 + c_{20} + c_{23} \Big) (P^2)^{\mu\nu} P_{\nu aA} \Big]
\nn\\
& + \tblue{( \EA \g_{\m\n} D_\rh \psi^a )}
\Big[\Big( \tfrac12 c_5 + c_{26} + c_{30} \Big)\,P^\rho{}_a{}^B Q^{\mu\nu}_{AB} 
\nn\w2
& + \Big( - \tfrac12 c_5 + c_3 + c_{25} + 2 c_{30} \Big) P^\mu{}_a{}^B Q^{\nu\rho}_{AB}  
\nn\\
& +\Big( - c_4 - \tfrac12 c_5 + 2 c_6 - c_{27} \Big) P^\m{_{aA}} (P^2)^{\n\rh}
- \tfrac12 c_{28}  P_{\s a A} R^{\m\n \rh \s} \Big]\ .
\label{V1}
\end{align}
%
% 
%%%%%%%%%%%%%%%%%%%%%%%%%%%%%%%%%%%%%%%%%%%%%%%%%%%%%%%%%%%%%%%%%%%%%
% \subsection{Dilatino  terms}
%%%%%%%%%%%%%%%%%%%%%%%%%%%%%%%%%%%%%%%%%%%%%%%%%%%%%%%%%%%%%%%%%%%%%
%
Next, we collect all the variations involving the dilatino. They are given by
\begin{align}
V(\chi) &= \tblue{\big(\eb\chi\big)} \Big[ \big( - 2 \al + 2 b_1 - c_1 + \tfrac12 c_2 + \tfrac14 c_{16} + \tfrac12 c_{22} \big) Q^2
\nn\\
& +\big( 2 b_2  - \tfrac12 c_{17} - c_{23} \big) (P^2)_{\mu\nu} (P^2)^{\mu\nu} 
+\big( 2 b_3  - \tfrac12 c_{18} - c_{24} \big) (P^2)^2 - c_{15} \Box P^2 \Big]
\nn\\
&+\tblue{( \EA \g^{\m\n} \chi^B )} \Big[ \big( - c_8 - 4 c_{13} \big)\, ( D_\m P^{\rh a}{_A} ) ( D_\n P_{\rh a B} )
\nn\\
& + \big( \tfrac14 c_8 + \tfrac12 c_{16} + c_{29} + c_{22} - 2 c_{30} - 3 c_{13} \big)  \, \big( Q_{\mu\lambda} Q_\nu{}^\lambda \big)_{AB}  
\nn\\
& + \big( \tfrac12 c_8 - \tfrac12 c_{16} + \tfrac12 c_{17} + c_{29} - c_{22} + c_{23} - 2 c_{30} + 6 c_{13} \big) \,(P^2)_\mu{}^\s Q_{\nu\s AB}
\nn\\
& 
+\big( - \tfrac14 c_8 - \tfrac12 c_{18} + \tfrac12 c_{29} - c_{24} - c_{30} - c_{13} \big) \, Q_{\m\n AB} P^2 
\nn\\
& + \big(  - \tfrac12 c_1 + \tfrac14 c_2 - \tfrac14 c_8 \big) R_{\m\n}{}^{\rho\sigma} Q_{\rho\sigma AB}\, \Big] 
\nn\\
& + \big(  - \tfrac14 c_{29} + \tfrac12 c_{30} + \tfrac14 \g \big) \tblue{( \bar{\e} \g^{\m\n\rh\s} \chi )} \big(Q_{\m\n} Q_{\rho\sigma}\big)  
\nn\\
& +\big( 4 c_7 - c_8 - c_9 \big) \,\tblue{( \EA D^\m \chi^B )} (PDP)_{\m AB}
+ \big( - c_{15} + 2 c_{14} \big)\tblue{(\eb D^\m \chi)} \pd_\m P^2 
\nn\\
& - c_{13} \tblue{( \EA \g^{\m\n} D^\rh \chi^B )}  \, D_\rho Q_{\m\n AB} - c_{11} \,\tblue{(\eb \g^{\m\n} D^\rh \chi)}  (PDP)_{\rh,\m\n} \ ,
\label{V2}
\end{align}
where we have used \eq{bp}. 

Finally, we turn to all the variations that contain the gravitino. They are given by
%%%%%%%%%%%%%%%%%%%%%%%%%%%%%%%%%%%%%%%%%%%%%%%%%%%%%%%%%%%%%%%%%%%%%%%%%%%%%%%%
%\subsection{Gravitino terms}
%%%%%%%%%%%%%%%%%%%%%%%%%%%%%%%%%%%%%%%%%%%%%%%%%%%%%%%%%%%%%%%%%%%%%%%%%%%%%%%%%
%
\begin{align}
V(\psi_\m)
&= \tblue{(\EA \g_\m \psi^B_\n)} \Big[ \big( - 4 \al - 2 c_1 - c_3 - \tfrac12 c_{19} - \tfrac12 c_{25} - c_{26} + \tfrac12 c_{22} - 2 c_{30} \big) \big(Q^\mu{}_\s Q^{\nu\s}\big)_{(AB)} 
\nn\\
& + \big( 4 \al + 2 c_1 + \tfrac12 c_{19} - \tfrac12 c_{25} - \tfrac12 c_{22} - c_{30} \big) (P^2)^{\m\s} Q^\n{}_{\s AB} 
\nn\\
& + \big( - c_4 - \tfrac12 c_{20} + c_{26} - \tfrac12 c_{27} - \tfrac12 c_{23} + c_{30} \big) (P^2)^{\n\s} Q^\m{}_{\s AB}
\nn\\
& + \big( - \tfrac12 c_3 - \tfrac12 c_{21} - \tfrac12 c_{25} - \tfrac12 c_{24} - c_{30} \big) Q^{\m\n}{}_{AB} P^2
\nn\\
& + \tfrac14 c_{28}  Q_{\rh\s AB} R^{\m\n\rh\s} \Big]
\nn\\
& + \tblue{( \eb \gamma^\mu \psi^\nu )} \Big[ \big( 2 \al - 4 b_1 + c_1 - \tfrac12 c_3 + \tfrac34 c_{10} - \tfrac14 c_{19} - \tfrac14 c_{25} + \tfrac12 c_{26} - \tfrac34 c_{22} + \tfrac32 c_{14} \big) \big(Q_{\mu\rho} Q_\nu{}^\rho\big)
\nn\\
& + \big( - 4 b_2  + \tfrac32 c_{10} - \tfrac12 c_{20} + \tfrac12 c_{27} + \tfrac32 c_{23} - c_{14} \big) (P^2)_{\mu\rho} (P^2)_\nu{}^\rho 
\nn\\
& +\big( \tfrac12 c_4 - 4 b_3  + \tfrac12 c_{10} - \tfrac12 c_{21} - \tfrac12 c_{27} + \tfrac32 c_{24} + c_{14} \big) (P^2)_{\mu\nu} P^2 
\nn\\
& + \big( - c_4 + \tfrac12 c_{28} + 2 c_{14}\big) \, (P^2)^{\rho\sigma} R_{\mu\rho\nu\sigma} 
\nn\\
& + \big( c_{10} + 2 c_{14} \big) \left(D_\m P_{\rh aA} \right)\left(D_\n P^{\rh aA}\right) + \big( - c_{10} + 2 c_{14} \big) P^{\rh\, aA} D_\rh D_\m P_{\n aA} \Big] 
\nn\\
&
+ \tblue{( {\bar\e} \gamma^\mu \psi_\mu )} \Big[ \big( - \al + b_1 - c_1 + \tfrac12 c_{22} \big) Q^2 
+ \big( b_2 - c_{23} \big)  (P^2)_{\nu\rho} (P^2)^{\nu\rho}
\nn\\
& + \big( b_3 - c_{24} \big) (P^2)^2 
- c_{14} \Box P^2 \Big] 
\nn\\
& + \tblue{( {\bar\e}^A \g_{\n\rh\s} \psi_\m^B )} \Big[ \big( - \tfrac12 c_3 - \tfrac12 c_{26} - \tfrac32 c_{30} \big) \big(Q^{\m\n} Q^{\rh\s} \big)_{AB} + \tfrac12 c_{25} \big( Q^{\m\n} Q^{\rh\s} \big)_{BA} 
\nn\\
& + \big( \tfrac12 c_4 - \tfrac12 c_{26} - \tfrac12 c_{27} + \tfrac32 c_{30} \big) (P^2)^{\m\n} Q^{\rh\s}{}_{AB} 
+ \big(   \tfrac12 c_3 + \tfrac14 c_{28} \big) R^{\ta\m\n\rh} Q^\s{}_{\ta AB} \Big] 
\nn\\
& + \tblue{( \EA \g^{\m\n\rh} \psi^B_\rh )} \Big[ - 2 c_7 \big( D_\m P^{\s a}{}_A \big) \big( D_\n P_{\s aB} \big) + \big( - \tfrac12 c_1 - \tfrac12 c_7 \big) R_{\m\n}{^{\s\ta}} Q_{\s\ta AB}
\nn\\
& + \big( - \tfrac12 c_7 - c_{24} - c_{30} \big) Q_{\m\n A B} P^2 + \big( c_7 - c_{22} + c_{23} - 2 c_{30} \big) (P^2)_\m{}^\s Q_{\n\s AB}
\nn\\
& + \big( \tfrac12 c_7 + c_{22} - 2 c_{30} \big) \big( Q_\m{}^\s Q_{\n\s} \big)_{AB} \Big] 
\nn\\
& - \tfrac14 c_{28} \tblue{( \eb \g_{\n\rh\s} \psi_\m )} R^{\m\tau \n\rh } (P^2)_\tau{}^\s + \big( \tfrac12 c_{30} + \tfrac18 \gamma \big) \tblue{(\eb \g^{\m\n\rh\s\ta} \psi_\m)} \big(Q_{\n\rh}{}Q_{\s\ta}\big)
\nn\\
& + \big( - c_1 - \tfrac12 c_3 \big) \tblue{( \EA \gamma^\m \psi_{\nu\rho}^B )} D_\m Q^{\nu\rho}_{AB} + \big(  - c_4 - c_{10} \big) \tblue{( \bar{\e} \g^\rh \psi^{\mu\nu} )} (PDP)_{\rh, \m\n}\ ,
\label{V3}
\end{align}
where we have used \eq{ddp} and \eq{dp12}. 

Requiring that the coefficients of all the structures listed above vanish, gives the following result:
\begin{align}
b_1 =&\, \al + \tfrac14 \g \ , 	&	b_2 =&\, - \g \ , 			            &	b_3 =&\, \tfrac14 \g     \ , 	        &    c_1=&\, 0 \ ,     &   c_2 =&\, 0 \ , 
\nn\w2
c_3 =&\, 0 \ , 					&	c_4 =&\, 0 \ , 			        &	c_5 =&\, - \g \ , 		        &	c_6 =&\, - \g \ ,      &	c_7 =&\, 0 \ , 			
\nn\w2
c_8 =&\, 0 \ , 			        &	c_9 =&\, 0 \ , 				&	c_{10} =&\, 0 \ , 				                &	c_{11} =&\, 0 \ , 				&	c_{12} =&\, \tfrac32 \g \ , 	
\nn\w2
c_{13} =&\, 0 \ ,     	    	&	c_{14} =&\, 0 \ , 			                   	&	c_{15} =&\, 0 \ ,           &   c_{16} =&\, - \g \ , 	&	c_{17} =&\, - 2 \g \ , 
\nn\w2
c_{18} =&\, \tfrac12 \g \ , 	&	c_{19} =&\, - 8 \al - \g \ ,   &   c_{20} =&\, \tfrac72 \g \ , 	&	c_{21} =&\, \tfrac14 \g \ , 	&	c_{22} =&\, - \tfrac12 \g \ , 
\nn\w2
c_{23} =&\, - \g \ , 	 &	c_{24} =&\, \tfrac14 \g \ , 	&	c_{25} =&\, 0 \ , 			 &	c_{26} =&\, \tfrac34 \g \ , 	&	c_{27} =&\, - \tfrac32 \g \ , 		
\nn\w2
c_{28} =&\, 0 \ ,   &	c_{29} =&\, \tfrac12 \g \ ,     &   c_{30} =&\, - \tfrac14 \g \ . 
\label{sol}
\end{align}

\subsection*{ The $\vp_\mu$ dependent variations }
%%%%%%%%%%%%%%%%%%%%%%%%%%%%%%%%%%%%%%%%%%%%%%%%%%%%%%%%%

So far we have considered the $H$ and $\vp_\mu$ independent variations of the action. Now that we have found the solution \eq{sol} for such variations to vanish, we shall now examine all such variations as well. In this subsection, we shall consider variations that contain at least a factor of $\vp_\mu$ but no $H$-dependence. Such variations arise from, (a)  the variation of the dilatino, (b) from integrations by part in which the exponential in dilaton factor is differentiated, and (c) the use of the EOM's. In the last case, we make use \eq{be} and \eq{fe}. Thus, without assuming the solution \eq{sol}, all the $\vp_\mu$ dependent variations are found to be
\begin{align}
V_{\partial\vp} =&\, \del \cL\Big|_{\partial\vp}
\nn\w2
=&  e e^{2\vp} \Bigg\{  \tblue{( \bar{\e}^A \psi^a )} \Big[ ( - \tfrac12 c_{18} + c_{24} ) P^\m{}_{a A} P^2 \vp_\m + ( \tfrac12 c_{16} - c_{22} ) P_{\n a}{}^B Q^{\m\n}{}_{AB} \vp_\m + ( - \tfrac12 c_{17} + c_{23} ) P_{\n a A} (P^2)^{\m\n} \vp_\m \Big] 
\nn\\
&\, + \tblue{( \bar{\e}^A \g_{\m\n} \psi^a )} \Big[ ( \tfrac12 c_{18} - c_{24} ) P^\m{}_{a A} P^2 \vp^\n + ( \tfrac12 c_{16} - c_{22} ) P_{\rh a}{}^B Q^{\n\rh}{}_{AB} \vp^\m + ( - \tfrac12 c_{17} + c_{23} ) P_{\rh a A} (P^2)^{\n\rh} \vp^\m 
\nn\\
&\, + ( \tfrac12 c_{29} + c_{30} ) P^\rh{}_a{}^B Q^{\m\n}{}_{AB} \vp_\rh + ( c_{29} + 2 c_{30} ) P^\m{}_a{}^B Q^{\n\rh}{}_{AB} \vp_\rh \Big] 
\nn\\
&\, + ( - \tfrac12 c_{29} - c_{30} ) \tblue{( \bar{\e}^A \g_{\m\n\rh\s} \psi^a )} P^\m{}_a{}^B Q^{\n\rh}{}_{AB} \vp^\s \Bigg\}\ ,
\end{align}
with $\cL$ from \eq{Lag}. Employing the solution \eq{sol}, we see that $V_{\partial\vp}=0$. Therefore, we do not need to add any new term to the Lagrangian \eq{Lag}\footnote{Terms such as $(P^2)^{\mu\nu} {\bar\chi}\gamma_\mu D_\nu \chi$ and $(PDP)^\m_{AB} {\bar\chi}^A \gamma_\m \chi^B$ contribute terms proportional to $\vp_\mu$, but such variations have different structures than those given in $V_\vp$ and they cannot be cancelled. Therefore the coefficients of such terms are vanishing in the solution we have found for the parameters in \eq{Lag}. }

\subsection*{The $H$-dependent variations}
%%%%%%%%%%%%%%%%%%%%%%%%%%%%%%%%%%%%%%%%%%%%%%%

Finally we consider all the remaining variations, namely those which contain at least a factor of $H$, or $H$ multiplied by  $\vp_\mu$ dependent factors. Collecting such variations, and omitting the EOM terms, we find that even though all variations involving more than one $H$ factor cancel each other, but terms linear in $H$ remain. To cancel all the $H$-dependent variations, we add the following terms to action:
\begin{equation}
\begin{aligned}
\cL_H = e\, e^{2 \vp} \big[&\,t_1 \bpsi^a \g^\m \psi^b Q^{\n\rh}{}_{ab} H_{\m\n\rh} + t_2 \bpsi^a \g^{\m\n\rh} \psi^b H_{\m\n}{}^\s (P^2)_{\rh\s ab} + t_3 \bpsi^a \g^{\m\n\rh} \psi^b H_{\m\n\rh} (P^2)_{ab} 
\w1
& + t_4 \bpsi^a \g^{\m\n\rh} \psi_a H_{\m\n}{}^\s (P^2)_{\rh\s} + t_5 \bpsi^a \g^{\m\n\rh} \psi_a H_{\m\n\rh} P^2 + t_6 \bpsi^a \g^{\m\n\rh\s\ta} \psi^b Q_{\m\n ab} H_{\rh\s\ta} \big]\ .
\end{aligned}
\end{equation}
The total contribution to $H$-depended variations, modulo EOM's, are then found to be
\begin{align}
V_H =&  \delta \Big( \cL +\cL_H \Big) \Big|_H  
\nn\w2
=& e e^{2\vp} \Bigg\{ ( 2 \g - \tfrac12 c_{26} - \tfrac12 c_{29} - \tfrac12 c_{30} + 2 t_1 ) \tblue{( \bar{\e}^A \psi^a )} P^\m{}_a{}^B Q^{\n\rh}{}_{AB} H_{\m\n\rh} 
\nn\\
&\, + \tblue{( \bar{\e}^A \g^{\m\n} \psi^a )} \Big[ ( 2 \al + \tfrac14 c_{16} + \tfrac14 c_{19} - \tfrac34 c_{22} - \tfrac12 t_2 + 3 t_3 ) P_{\s a}{}^B Q^{\rh\s}{}_{AB} H_{\m\n\rh}  
\nn\\
&\, + ( - \tfrac14 c_{17} + \tfrac14 c_{20} + \tfrac34 c_{23} + \tfrac12 t_2 + 3 t_3 + 2 t_4 ) P_{\s a A} (P^2)^{\rh\s} H_{\m\n\rh} 
\nn\\
&\, + ( \tfrac12 c_{29} + t_2 ) P_{\m a}{}^B Q^{\rh\s}{}_{AB} H_{\n\rh\s} 
\nn\\
&\, + ( c_{26} + c_{29} + 2 t_1 - t_2 ) P^\rh{}_a{}^B Q^\s{}_{\m AB} H_{\n\rh\s} 
\nn\\
&\, + ( \tfrac12 c_{27} - 2 t_1 - t_2 + 4 t_4 ) P^\rh{}_{a A} (P^2)^\s{}_\m H_{\n\rh\s} 
\nn\\
&\, + ( - \tfrac14 c_{18} + \tfrac14 c_{21} + \tfrac34 c_{24} + \tfrac12 t_2 + 6 t_5 )  P^\rh{}_{a A} P^2 H_{\m\n\rh} \Big]  
\nn\\
&\, + \tblue{( \bar{\e}^A \g^{\m\n\rh\s} \psi^a )} \Big[ ( \tfrac12 c_{29} + \tfrac32 c_{30} - \tfrac12 t_2 + 6 t_6 ) P_{\m a}{}^B Q_\n{}^\la{}_{AB} H_{\rh\s\la} 
\nn\\
&\, + ( \tfrac14 c_{26} + \tfrac14 c_{29} + \tfrac34 c_{30} + \tfrac12 t_2 ) P^\la{}_a{}^B Q_{\m\n AB} H_{\rh\s\la} 
\nn\\
&\, + ( \tfrac1{12} c_{16} - \tfrac16 c_{22} - t_3 - 2 t_6 ) P^\la{}_a{}^B Q_{\s\la AB} H_{\m\n\rh} 
\nn\\
&\, + ( - \tfrac14 c_{27} - \tfrac12 t_2 + 2 t_4 - 6 t_6 ) P_{\m a A} (P^2)_\n{}^\la H_{\rh\s\la} 
\nn\\
&\, + ( - \tfrac1{12} c_{17} + \tfrac16 c_{23} - t_3 + 2 t_6 ) P^\la{}_{a A} (P^2)_{\s\la} H_{\m\n\rh} 
\nn\\
&\, + ( \tfrac1{12} c_{18} - \tfrac16 c_{24} + 2 t_5 + 2 t_6 ) P_{\m a A} P^2 H_{\n\rh\s} \Big] 
\nn\\
&\, + ( - \tfrac1{12} c_{29} - \tfrac16 c_{30} + 2 t_6 ) \tblue{( \bar{\e}^A \g^{\m\n\rh\s\la\ta} \psi^a )} P_{\m a}{}^B Q_{\n\rh AB} H_{\s\la\ta} \Bigg\} \ ,
\label{TH}
\end{align}
with $\cL$ from \eq{Lag}. Upon taking the following values
\begin{equation}
t_1 = - \tfrac34 \g \ , \qquad t_2 = t_4 = - \tfrac14 \g \ , \qquad t_3 = t_5 = t_6 = 0 \ , 
\label{Hsol}
\end{equation}
and using \eqref{sol} and \eqref{Hsol}, we get
\begin{equation}
V_H = 0 \ . 
\end{equation}
Thus the invariant total Lagrangian is
\be
\cL=\cL(R) + \cL(P^2) + \beta \cL(F^2) +  \alpha \cL(R^2) +\cL_{\alpha,\gamma} (P^4) 
+ \cL_H\ ,
\label{TLag}
\ee
with parameters given in \eq{sol} and \eq{Hsol}. In particular, $\cL_H$ is given by
\be
\cL_H = \g\, e\, e^{2 \vp} \big[ - \tfrac34 \bpsi^a \g^\m \psi^b Q^{\n\rh}{}_{ab} H_{\m\n\rh} - \tfrac14 \bpsi^a \g^{\m\n\rh} \psi^b H_{\m\n}{}^\s (P^2)_{\rh\s ab} - \tfrac14 \bpsi^a \g^{\m\n\rh} \psi_a H_{\m\n}{}^\s (P^2)_{\rh\s} \big]\ .
\label{LH}
\ee
The last two terms can be interpreted as bosonic torsion in the $c_5$ and $c_6$ terms in $\cL_2$ (up to quartic fermions). To be more specific, we have
\begin{align}
- \tblue{( \bpsi^a \g_\m D_\n \psi^b )} (P^2)^{\m\n}{}_{ab} - \tfrac14 \tblue{( \bpsi^a \g^{\m\n\rh} \psi^b )} (P^2)_{\rh\s ab} H_{\m\n}{}^\s =&\, - \tblue{( \bpsi^a \g_\m D_\n(\Omega_+) \psi^b )} (P^2)^{\m\n}{}_{ab} \ ,
\nn\w2
- \tblue{( \bpsi^a \g_\m D_\n \psi_a )} (P^2)^{\m\n} - \tfrac14 \tblue{( \bpsi^a \g^{\m\n\rh} \psi_a )} (P^2)_{\rh\s} H_{\m\n}{}^\s =&\, - \tblue{( \bpsi^a \g_\m D_\n(\Omega_+) \psi_a )} (P^2)^{\m\n}\ .
\end{align}
The first term in \eq{LH} has no torsion interpretation directly. The explicit expression for \eq{TLag} is given below in \eq{FL}. 

%%%%%%%%%%%%%%%%%%%%%%%%%%%%%%%%%%%%%%%%%%%%%%%%%%%%%%%%%%%%%%
\subsection{EOM terms and supertransformations}
%%%%%%%%%%%%%%%%%%%%%%%%%%%%%%%%%%%%%%%%%%%%%%%%%%%%%%%%%%%%%%

We shall now determine all the EOM terms we have suppressed in the supersymmetry variations so far. These are the terms which dropped in \eq{V1}, \eq{V2}, \eq{V3} and \eq{TH}. This needs to be done so that we can determine the field redefinitions required to cancel them, and the consequences for the supertransformations. Collecting all the EOM terms that arise in the variation of the total Lagrangian \eq{TLag} given above, we find 
\begin{align}
V_{\rm EOM} =&\, \del \cL|_{EOM}
\nn\w2
=&\, \al\, e\, e^{2 \vp} \Big[ 8 \tblue{( \bar{\e}^A \g_\m \psi_\n^B )} Q^{\n\rh}{}_{AB} \EW^\m{}_\rh + 2 \tblue{( \bar{\e}^A \g_\m \psi_\n^B )} Q^\n{}_{\rh AB} e^{- 2 \vp} \mathcal{E}_B^{\m\rh} 
\nn\\
&\, - 2 \tblue{( \bar{\e}^A D_\m \E_\n^B )} Q^{\m\n}{}_{AB} + \tfrac12 \tblue{( \bar{\e}^A \g_{\m\n} \E_\rh^B )} Q^{\rh\s}{}_{AB} H^{\m\n}{}_\s 
\nn\w1
&\, - 8 \tblue{( \bar{\e}^A \g_\m D_\n \E^B )} Q^{\m\n}{}_{AB} - 4 \tblue{( \bar{\e}^A \g_\m \E^B )} Q_{\n\rh AB} H^{\m\n\rh} 
\nn\\
&\, - 2 \tblue{( \bar{\e}^A \g^{\m\n\rh} \E^B )} Q_\rh{}^\s{}_{AB} H_{\m\n\s} \Big] 
\nn\\
&\, + \g\, e\, e^{2 \vp} \Big[ \tblue{( \bar{\e}^A \psi^a )} ( - 3 P_{\m a (A|} P^\m{}_{b |B)} \EW^{b B} + (P^2)_{ab} \EW^b{}_A + P^2 \EW_{a A} ) 
\nn\\
&\, + \tblue{( \bar{\e}^A \g_{\m\n} \psi^a )} ( P^\m{}_{a (A|} P^\n{}_{b |B)} \EW^{b B} + \tfrac12 Q^{\m\n}{}_{AB} \EW_a{}^B ) 
\nn\w1
&\, + \tblue{( \bar{\e}^A \g^\m \E^a )} ( - \tfrac12 P^\n{}_a{}^B Q_{\m\n AB} + P^\n{}_{a A} (P^2)_{\m\n} - \tfrac14 P_{\m a A} P^2 ) 
\nn\\
&\, + \tfrac14 \tblue{( \bar{\e}^A \g^{\m\n\rh} \E^a )} P_{\rh a}{}^B Q_{\m\n AB} ) \Big] \ .
\label{eomv}
\end{align}
 We shall now cancel these by modifying the SUSY transformation rules. Denoting the modification of supertransformation rules by $\del_{\rm extra}$,  we get the following extra terms in the variation of Lagrangian. 
\begin{align}
\del_{\rm extra} \cL = e\, e^{2 \vp} \Big[&\, - 2 e_{\n r} ( \del_{\rm extra} e_\m{}^r ) \E^{\m\n} + ( \del_{\rm extra} \vp ) \E_{\vp} + \tfrac12 ( \del_{\rm extra} B_{\m\n} ) e^{- 2 \vp} \mathcal{E}_B^{\m\n} + V_{\al a A} ( \del_{\rm extra} \phi^\al ) \EW^{a A} 
\nn\w1
&\, - \tfrac12 \bar{\E}^{\m A} ( \del_{\rm extra} \psi_{\m A} ) + 4 \bar{\E}^A ( \del_{\rm extra} \chi_A ) - \bar{\E}^a ( \del_{\rm extra} \psi_a ) \Big] \ .
\label{delex}
\end{align}
Requiring that these variations cancel $V_{EOM}$ given above, we find that the supertransformations \eq{super2} need to be supplemented by $\delta_{extra}$ given by \footnote{Note that integration by part in the two terms that have the form $D_\m\E^B  Q^{\m\n}_{AB}$ gives terms of the form $\E_\n^B \E_{aA}$. This means that we can alternatively redefine the hyperscalar to remove such variation.} 
\begin{align}
\del_{\rm extra} e_\m{}^r =&\, - 4 \al ( \bar{\e}^A \g^r \psi_\n^B ) Q_\m{}^\n{}_{AB} \ , 
\nn\w2
\del_{\rm extra} \psi_{\m A} =&\, 4 \al ( \del_0 \psi_\n^B ) Q_\m{}^\n{}_{AB} - 8 \al \e^B (PDP)_{\m AB} + 8 \al \e^B P_\m{}^a{}_{(A|} \EW_{a |B)} \ , 
\nn\w2
\del_{\rm extra} B_{\m\n} =&\, 4 \al ( \bar{\e}^A \g_{[\m|} \psi_\rh^B ) Q_{|\n]}{}^\rh{}_{AB} \ , 
\nn\w2
\del_{\rm extra} \chi_A =&\, - 2 \al \g^\m ( \del_0 \psi_\n^B ) Q_\m{}^\n{}_{AB} + 4 \al \g^\m \e^B (PDP)_{\m AB} - 4 \al \g^\m \e^B P_\m{}^a{}_{(A|} \EW_{a |B)} \ , 
\nn\w2
\del_{\rm extra} \vp =&\, 2 \al ( \bar{\e}^A \g_\m \psi_\n^B ) Q^{\m\n}{}_{AB} \ , 
\nn\w2
V_{\al a A} \del_{\rm extra} \phi^\al =&\, 3 \g ( \bar{\e}^B \psi^b ) P^\m{}_{a (A|} P_{\m b |B)} - \g ( \bar{\e}_A \psi^b ) (P^2)_{ab} - \g ( \bar{\e}_A \psi_a ) P^2 
\nn\\
&\, + \g ( \bar{\e}^B \g_{\m\n} \psi^b ) P^\m{}_{a (A|} P^\n{}_{b |B)} + \tfrac12 \g ( \bar{\e}^B \g_{\m\n} \psi_a ) Q^{\m\n}{}_{AB} \ , 
\nn\w2
\del_{\rm extra} \psi_a =&\, \g ( \g^\m \e^A ) P^\n{}_{a A} (P^2)_{\m\n} - \tfrac14 \g ( \g^\m \e^A ) P_{\m a A} P^2 - \tfrac14 \g ( \g^{\m\n} \g^\rh \e^A ) P_{\rh a}{}^B Q_{\m\n AB} \ , 
\end{align}
where $\delta_0$ is the supersymmetry variation which is zeroth order in $\alpha,\beta$ and $\gamma$. The $\psi^\n Q_{\m\n}$ and $(\delta_0 \psi^\n ) Q_{\m\n}$ terms in the first five transformation rules above can be removed by the field redefinitions 
\be
\psi_{\m A} \to \psi_{\m A}+4\alpha\,\psi^{\nu B} Q_{\m\n AB} \ ,
\qquad \chi_A \to \ \chi_A -2\alpha\, \gamma_\m \psi_\n^B Q^{\m\n}{}_{AB} \ .
\label{fr}
\ee
Consequently, this redefinition bring in terms of the form $\alpha \bar{\E}^\m \psi^\n Q_{\m\n}$ and $\alpha \bar{\E} \g^\m \psi^\n Q_{\m\n}$ in the Lagrangian, which are presented in the final results summarized in the next section.

\section{The final results}
%%%%%%%%%%%%%%%%%%%%%%%%%%%%%%%%%%%%%%%%%%%%%%%%%%%%%%%%%%%%%%%%%%%%%%%%%%%%%%%%%%%%%

In summary, substituting the parameter values given in \eq{sol} into the Lagrangian $\cL_{\alpha,\gamma}$ given in \eq{L3c}, adding the Lagrangian $\cL_H$ given in \eq{LH}, and performing the field redefinition \eq{fr}, our result for the total Lagrangian \eq{TLag} is given by 
\begin{align}
\cL = e\, e^{2 \vp} \Big[&\, \tfrac{1}{4} R(\omega) + \vp_\m \vp^\m - \tfrac1{12} \cH_{\m\n\rh} \cH^{\m\n\rh} - \tfrac12 \bpsi_\m \g^{\m\n\rh} D_\n \psi_\rh + 2 \bar{\chi} \g^{\m\n} D_\m \psi_\n + 2 \bar{\chi} \g^\m D_\m \chi 
\nn\w2
&\, - \tfrac1{24} \cH_{\m\n\rh} \left( \bpsi^\s \g_{[\s} \g^{\m\n\rh} \gamma_{\ta]} \psi^\ta + 4 \bpsi_\s \g^{\s\m\n\rh} \chi - 4 \bar{\chi} \g^{\m\n\rh} \chi \right) - \vp_\m \left( \bpsi^\m \gamma^\n \psi_\n + 2 \bpsi_\n \g^\m \g^\n \chi \right) 
\nn\w2
&\, -\tfrac12 P^{\m a A} P_{\m a A} - \tfrac12 \bpsi^a \g^\m D_\m \psi_a - \tfrac1{24} \cH_{\m\n\rh} \bpsi^a \g^{\m\n\rh} \psi_a - P_{\m a A} \left( \bpsi_\n^A \g^\m \g^\n \psi^a + 2 \bar{\chi}^A \g^\m \psi^a \right) \Big] 
\nn\w2
+ \beta\, e\, e^{2\varphi} \Big[&\,
- \tfrac{1}{4} F_{\mu\nu}^I F^{I\mu\nu}
- \bar{\lambda}^I \gamma^\mu D_\mu \lambda^I - \tfrac{1}{12} \cH_{\mu\nu\rho} 
\bar{\lambda}^I \gamma^{\mu\nu\rho} \lambda^I
+ \tfrac{1}{2} F_{\mu\nu}^I \bar{\lambda}^I \left( 
\gamma^\rho \gamma^{\mu\nu} \psi_\rho + 2 \gamma^{\mu\nu} \chi \right) \Big] 
\nn\w2
+ \al\, e\, e^{2 \vp} \Big[&\,- \tfrac14 R_{\m\n}{}^{rs}(\Omega_-) R^{\m\n}{}_{rs}(\Omega_-) + Q^{\m\n AB} Q_{\m\n AB} - \bar{\psi}^{rs} \slashed{D}(\omega, \Omega_-) \psi_{rs} 
\nn\w2
&\, + \tfrac12 \bar{\psi}_{rs} \big( \g^\ta \g^{\m\n} \psi_\ta + 2 \g^{\m\n} \chi \big) R_{\m\n}{}^{rs}(\Omega_-) - \tfrac1{12} \cH_{\m\n\rh} \bar{\psi}^{rs} \g^{\m\n\rh} \psi_{rs} - 8 \bpsi_\m^A \psi_a Q^{\m\n}{}_{AB} P_\n{}^{a B} 
\nn\w2
&\, + 2 \bpsi_\n^B \E_\m^A Q^{\m\n}{}_{AB} - 8 \bpsi_\n^B \g_\m \E^A Q^{\m\n}{}_{AB} \Big] 
\nn\w2
+ \g\, e\, e^{2 \vp} \Big[&\, \tfrac14 Q^{\m\n AB} Q_{\m\n AB} - (P^2)^{\m\n} (P^2)_{\m\n} + \tfrac14 (P^2)^2 
\nn\w2
&\, -( \bpsi^a \g_\m D_\n(\Omega_+) \psi^b ) (P^2)^{\m\n}{}_{ab} 
 - (\bpsi^a \g_\m D_\n(\Omega_+) \psi_a ) (P^2)^{\m\n} - \tfrac34 \bpsi^a \g^\m \psi^b Q^{\n\rh}{}_{ab} \cH_{\m\n\rh} 
\nn\w2
&\, + \tfrac32 \bpsi^a \g_\m \psi^b (PDP)^\m{}_{ab} + \tfrac12 \bar{\chi}^A \g^{\m\n\rh} \psi^a Q_{\m\n A}{}^B P_{\rh a B} - \tfrac14 \bpsi_\m^A \g^{\m\n\rh\s} \psi^a Q_{\n\rh A}{}^B P_{\s a B} 
\nn\w2
&\, + \bar{\chi}_A \g^\m \psi_a ( - Q_{\m\n}{}^{AB} P^{\n a}{}_B - 2 (P^2)_{\m\n} P^{\n a A} + \tfrac12 P_\m{}^{a A} P^2 ) 
\nn\w2
&\, + \bpsi_\m^A \psi_a ( - Q^{\m\n}{}_{AB} P_\n{}^{a B} + \tfrac72 (P^2)^{\m\n} P_\n{}^a{}_A + \tfrac14 P^{\m a}{}_A P^2 )  
\nn\w2
&\, + \bpsi_{\m A} \g^{\m\n} \psi^a ( - \tfrac12 Q_{\n\rh}{}^{AB} P^\rh{}_{a B} - (P^2)_{\n\rh} P^\rh{}_a{}^A + \tfrac14 P_{\n a}{}^A P^2 ) 
\nn\w2
&\, + \bpsi_\m^A \g_{\n\rh} \psi_a ( \tfrac34 Q^{\n\rh}{}_{AB} P^{\m a B} - \tfrac32 (P^2)^{\m\n} P^{\rh a}{}_A ) \Big] \ ,
\label{FL}
\end{align}
where $\cH_{\m\n\rh}$ is defined in \eq{CH}, and only terms up to first order in $\alpha, \beta$ and $\gamma$ are to be kept. There are two terms in the $\alpha$ dependent part of the Lagrangian which have  higher derivative hypermultiplet dependence. Schematically these terms are of the form $P^4$ and $\psi_\m \psi_a P^3$. The $\gamma$ dependent terms above, together with these two terms, constitute the Lagrangian $\cL_{\alpha,\gamma}$ appearing in \eq{Lag}.

Taking into account the field redefinitions \eq{fr}, the supertransformations are given by
\begin{align}
\del e_\m{}^r =&\, \bar{\e} \g^r \psi_\m \ , 
\nn\w2
\del \psi_{\m A} =&\, D_\m \e_A + \tfrac14 \cH_{\m\n\rh} \g^{\n\rh} \e_A - 8 \al \e^B (PDP)_{\m AB} + 8 \al \e^B P_\m{}^a{}_{(A|} \EW_{a |B)} \ , 
\nn\w2
\del B_{\m\n} =&\, - \bar{\e} \g_{[\m} \psi_{\n]} +2 \beta\, A_{[\m}^I \delta A_{\n]}^I + 2 \al\, \Omega_{- [\m}{}^{rs} \del_0 \Omega_{- \n] rs} + 2 \g\, Q_{[\m}{}^{AB} \del_0 Q_{\n] AB} \ , 
\nn\w2
\del \chi_A =&\, \tfrac12 \g^\m \e_A \pd_\m \vp - \tfrac1{12} \cH_{\m\n\rh} \g^{\m\n\rh} \e_A + 4 \al \g^\m \e^B (PDP)_{\m AB} - 4 \al \g^\m \e^B P_\m{}^a{}_{(A|} \EW_{a |B)} \ , 
\nn\w2
\del \vp =&\, \bar{\e} \chi \ , 
\nn\w2
V_{\al a A} \del \phi^\al =&\, - \bar{\e}_A \psi_a + 3 \g\, \bar{\e}^B \psi^b P^\m{}_{a (A|} P_{\m b |B)} - \g\, \bar{\e}_A \psi^b (P^2)_{ab} - \g\, \bar{\e}_A \psi_a P^2 
\nn\w2
&\, + \g\, \bar{\e}^B \g_{\m\n} \psi^b P^\m{}_{a (A|} P^\n{}_{b |B)} + \tfrac12 \g\, \bar{\e}^B \g_{\m\n} \psi_a Q^{\m\n}{}_{AB} \ , 
\nn\w2
\del \psi_a =&\, \g^\m \e^A P_{\m a A} + \g ( \g^\m \e^A ) P^\n{}_{a A} (P^2)_{\m\n} 
\nn\w2
&\, - \tfrac14 \g ( \g^\m \e^A ) P_{\m a A} P^2 - \tfrac14 \g ( \g^{\m\n} \g^\rh \e^A ) P_{\rh a}{}^B Q_{\m\n AB} \ , 
\nn\w2
\delta A_\mu^I =&\, - \bar{\epsilon} \gamma_\mu \lambda^I \ , 
\nn\w2
\delta \lambda^I
=&\, \tfrac{1}{4} F_{\mu\nu}^I \gamma^{\mu\nu} \epsilon \ . 
\label{susy1}
\end{align}

\section{$Hp(1)$ from $Hp(n)$ compared with heterotic supergravity on $T^4$} 
%%%%%%%%%%%%%%%%%%%%%%%%%%%%%%%%%%%%%%%%%%%%%%%%%%%%%%%%%%%%%%%%%%%%%%%%%%%%%%%%%

It has been shown that the dimensional reduction of the BdR extended heterotic supergravity on $T^4$ followed by a truncation to $N=(1,0)$ supersymmetry yields the higher derivative couplings of four hypermultiplets which parametrize $SO(4,4)/SO(4)\times SO(4)$ \cite{Eloy:2020dko, Chang:2021tsj}. Truncating further to keep  a single hypermultiplet give the coset $SO(4,1)/SO(4)$. Given that $SO(4,1) \approx Sp(1,1)$ and $SO(4) \approx Sp(1)\times Sp(1)$, we can compare our results for $Hp(n)$ model truncated to $Hp(1)$. To do so, we begin with the truncation of the $Hp(n)$ model to obtain the $Hp(1)$ model.

\subsection{The $Hp(1)$ model from the truncation of the $Hp(n)$ model}
%%%%%%%%%%%%%%%%%%%%%%%%%%%%%%%%%%%%%%%%%%%%%%%%%%%%%%%%%%%%%%%%%%%%%%%%%

To truncate the $Hp(n)$ model to $Hp(1)$, we let the index $a=1,2$. This implies the identities
\bea
(P^2)^{\m\n}{}_{ab} &=& - \tfrac12 (P^2)^{\m\n} \e_{ab} \ , 
\nn\w2
Q^2 &=&\, - 2 (P^2)_{\m\n} (P^2)^{\m\n} + 2 (P^2)^2\ ,
\nn\w2
P_{\n a}{}^B Q^{\m\n}{}_{AB} &=&\, - P_{\n a A} (P^2)^{\m\n} + P^\m{}_{aA} P^2 \ ,
\nn\w2
P^\m{}_a{}^B Q^{\n\rh}{}_{AB} \Big|_{[\m\n]} &=&\, P^\rh{}_a{}^B Q^{\m\n}{}_{AB} - 3 P^\m{}_{aA} (P^2)^{\n\rh} \ .
\label{HPL}
\eea
Using these identities in the final Lagrangian \eq{FL} in terms that correspond to those with coefficients $(b_1, c_5, c_{16}, c_{19}, c_{22}, c_{25})$ in \eq{L3c}, and leaving out $\beta \cL(F^2)$, and collecting the bosonic terms that are linear in $\alpha$ and $\gamma$, gives
\begin{align}
\cL_{\rm Bos.}\Big|_\al  =&\, e\,e^{2\varphi} \biggl[  H^{\m\n\rh} \big( \al\, \omega^L_{\m\n\rh}(\Omega_-)  + \gamma\, \omega^Q_{\m\n\rh} \big) - \tfrac14 \al\,R_{\m\n rs}(\Omega_-) R^{\m\n rs}(\Omega_-)  
\nn\w2
& -(2\alpha +\tfrac32 \gamma) (P^2)_{\mu\nu} (P^2)^{\mu\nu} + (2\alpha+\tfrac34\gamma) (P^2)^2 \biggr]\ ,
\label{HB}
\end{align}
where $\omega^Q_{\m\n\rho}$ is as defined in \eq{CSQ}.

\subsection{ The $Hp(1)$ from the dimensional reduction}
%%%%%%%%%%%%%%%%%%%%%%%%%%%%%%%%%%%%%%%%%%%%%%%%%%%%%%%%%%%%%%%%%%%%%%%%

As mentioned above, in the reduction of heterotic $10D$ supergravity on $T^4$, upon truncation to $N=(1,0)$, the resulting hyperscalars parametrize the coset $SO(4,4)/SO(4)_+\times SO(4)_-$, and it is important to note that the first index. Following the notation of \cite{Chang:2021tsj}, denoting a representative of this coset by $W$, we have the Maurer-Cartan form
\begin{equation}
W \partial_\mu W^{-1} = \left( 
\begin{array}{cc}
Q_{+ \mu ab} & \ \ - P_{\mu ab} \\
- P_{\mu ba} & \ \ Q_{- \mu ab}
\end{array}
\right)\ , 
\end{equation}
where $a,b=1,...,4$ and $Q_{\pm ab}$ are the $SO(4)_\pm$ connections. It is important to note that $P_{\m ab}$ transforms under $SO(4)_\pm$ as
\be
\delta P_{\m ab} = \Lambda_{+a}{}^c P_{\m\,cb} + \Lambda_{-b}{}^c P_{\m\,ac}\ .
\label{gt}
\ee
The $R$-symmetry group $Sp(1)_R \subset SO(4)_+$ \cite{Chang:2021tsj}. The bosonic part of the Lagrangian takes the form \cite{Eloy:2020dko} 
\begin{align}
\cL_{Bos. , \mathcal{O}(\al')} = \al e\, e^{2 \vp} \Big[&\, H^{\m\n\rh} \big( \omega^L_{\m\n\rh}(\Omega_-) - \omega^Q_{\m\n\rh}(Q_+) \big) - \tfrac14 R_{\m\n rs}(\Omega_-) R^{\m\n rs}(\Omega_-) 
\nn\w1
&\, + \tfrac12 \tr( P_\m P_\n^T ) \tr( P^\m P^{\n T} ) - \tfrac12 \tr( P_\m P_\n^T P^\m P^{\n T} ) 
\nn\w1
&\, + \tfrac12 \tr( P_\m^T P^\m P_\n^T P^\n ) - \tfrac12 \tr( P^\m P_\m^T P^\n P_\n^T ) \Big] \ , 
\label{RL}
\end{align}
with $\omega^L_{\m\n\rho}$ from \eq{CSQ} and 
\begin{align}
\omega^Q_{\m\n\rh}(Q_+) =&\, \Big(Q_{+ [\m a}{}^b \pd_\n Q_{+\rh b}{}^a + \tfrac23 Q_{+ \m a}{}^b Q_{+\n b}{}^c Q_{+ \rho c}{}^a \Big) \Big|_{[\m\n\rho]}
\nn\w2
=& 2 \Big( Q_{\m A}{}^B \partial_\n Q_{\rho B}{}^A +\tfrac23 Q_{\m A}{}^B Q_{\nu B}{}^C Q_{\rho C}{}^A \Big)_{[\m\n\rho]}
\nn\w2
&+2 \Big( Q_{\m A'}{}^{B'} \partial_\n Q_{\rho B'}{}^{A'} +\tfrac23 Q_{\m A'}{}^{B'} Q_{\nu B'}{}^{C'} Q_{\rho C'}{}^{A'} \Big)_{[\m\n\rho]}\ .
\end{align}
In obtaining the Lagrangian above, terms proportional to EOMs have been dropped on the basis that they can be removed by field redefinitions, as usual.

\subsection*{ Truncation with  $Sp(1) \subset SO(4)_+ $  }
%%%%%%%%%%%%%%%%%%%%%%%%%%%%%%%%%%%%%%%%%%%%%%%%%%%%%%%%%%%

Considering the truncated coset $SO(4,1)/SO(4)$ as locally being the same $Sp(1,1)/Sp(1)\times Sp(1)_R$, we first consider the truncation scheme in which the $Sp(1)$ factor here is embedded in the $SO(4)_+$. This amounts to letting the $SO(4)_-$ index to take one value, say $a=1$. Recalling that the second index on $P_{ab}$ is the $SO(4)_-$ index, we thus let 
\bea
&& P_{\m ab} \to P_{\m a 1} \equiv P_{\m a} = i (\sigma_a)_{A A'} P_\m{}^{A' A}\ ,\qquad \psi_a \to \psi_1 \equiv \psi \ ,
\nn\w2
&& Q_{-\m ab} \to 0\ .
\eea
It is convenient to define 
\bea
Q_{\m A}{}^B &\equiv& \tfrac14 Q_{+ \m a b} ( \s^{a b} )_A{}^B \ , \qquad Q_{\m}{}^{A'}{}_{B'} \equiv - \tfrac14 Q_{+ \m a b} ( \bar{\s}^{a b} )^{A'}{}_{B'} \ . 
\eea
Making use of the above results in the Lagrangian \eq{RL}, and making the following field redefinition in $\cL_0$, 
\begin{equation}
B_{\m\n} \to B_{\m\n} - 4 \alpha\, \theta_{\m\n} \ , \text{ where } \ \ \pd_{[\m} \theta_{\n\rh]} \equiv \omega^Q_{\m\n\rh}(Q^{AB}) + \omega^Q_{\m\n\rh}(Q^{A'B'}) \ ,
\end{equation}
the Chern-Simons terms cancel, and the first order in $\alpha$ sector of the Lagrangian \eq{RL} becomes
\begin{equation}
\cL_{\rm Bos.}\Big|_\al = \al e\, e^{2 \vp} \Big[ H^{\m\n\rh} \omega^L_{\m\n\rh}(\Omega_-) - \tfrac14 R_{\m\n rs}(\Omega_-) R^{\m\n rs}(\Omega_-) + Q^2 \Big] \ .  
\end{equation}
Noting the second identity in \eq{HPL}, this result agrees with the bosonic part of the  Lagrangian for the $Hp(1)$ model given in \eq{HB}, provided that we set $\gamma = 0$.

\subsection*{Truncation with $Sp(1) \subset SO(4)_-$ }
%%%%%%%%%%%%%%%%%%%%%%%%%%%%%%%%%%%%%%%%%%%%%%%%%%%%%%%%%%%%%%%%%%%%%%%%

There is another way to truncate the $SO(4,4)/SO(4)_+\times SO(4)_-$ such that in the resulting coset the surviving $Sp(1)$ factor is now embedded into $SO(4)_-$, instead of $SO(4)_+$ considered above. To see how this works, let us introduce the notation for the $SO(4)_+ \times SO(4)_-$ indices as follows,
\be
SO(4)_+ :\, \quad a \to A, A' \ ,\qquad\quad 
SO(4)_- :\, \quad a \to \bar{A}, \bar{A}' \ .
\ee
Thus we have,
\begin{align}
P_{\m a b} \equiv &\, \tfrac1{\sqrt{2}} ( \s_a )_{A A'} ( \s_b )_{\bar{A} \bar{A}'} P_\m{}^{A' A \bar{A}' \bar{A}} \ , 
&
\psi^{A'}{}_{\bar{A} \bar{A}'} \equiv&\, \tfrac{i}{\sqrt{2}} ( \s^a )_{\bar{A} \bar{A}'} \psi^{A'}_a \ , 
\nn\w2
Q_{ \m AB} \equiv&\, \tfrac14 Q_{+ \m a b} ( \s^{ab} )_{AB} \ , 
&
Q_{ \m \bar{A} \bar{B}} \equiv&\, \tfrac14 Q_{- \m a b} ( \s^{ab} )_{\bar{A} \bar{B}} \ , 
\nn\w2
Q_{ \m A'B'} \equiv&\, - \tfrac14 Q_{+ \m a b} ( \bar{\s}^{ab} )_{A'B'} \ , 
&
Q_{ \m}{}_{\bar{A}'\bar{B}'} \equiv&\, - \tfrac14 Q_{- \m a b} ( \bar{\s}^{ab} )_{\bar{A}'\bar{B}'} \ . 
\end{align}
The truncation such that $Sp(1)\subset SO(4)_-$ is implemented by setting
\bea
Q_{\m A'B'} &=& 0 = Q_{\m \bar{A}'\bar{B}'} \ ,\qquad P_\m{}^{1' A \bar{2}' \bar{A}} = 0 = P_\m{}^{2' A \bar{1}' \bar{A}}\ ,  \qquad \psi^{1' \bar{A} \bar{2}'} = 0 =\psi^{2' \bar{A} \bar{1}'}\ ,
\nn\w2
P_\m{}^{1' A \bar{1}' \bar{A}} &=&\, P_\m{}^{2' A \bar{2}' \bar{A}} \equiv - \tfrac1{\sqrt{2}} P_\m{}^{\bar{A} A} \ ,\qquad\psi^{1' \bar{A} \bar{1}'} =\, \psi^{2' \bar{A} \bar{2}'} \equiv \tfrac1{\sqrt{2}} \psi^{\bar{A}} \ .
\eea
The $Sp(1)_R$ connection $Q_{\m AB} \subset SO(4)_+$ as before, but now the $Sp(1)$ connection $Q_{\m \bar{A}\bar{B}} \subset SO(4)_-$. Performing the truncation described above in the bosonic part of the Lagrangian \eq{RL} gives
\begin{align}
\cL_{\rm Bos.}\Big|_\al = \al e\, e^{2 \vp} \Big[&\, H^{\m\n\rh} \big( \omega^L_{\m\n\rh}(\Omega_-) - 2 \omega_{\m\n\rh}^Q (Q^{AB})\big) 
\nn\w1
&\, - \tfrac14 R_{\m\n rs}(\Omega_-) R^{\m\n rs}(\Omega_-) + (P^2)_{\m\n} (P^2)^{\m\n} + \tfrac12 (P^2)^2 \Big] \ . 
\label{ae}
\end{align}
This results agrees with the bosonic part of our $Hp(1)$ model given in \eq{HB}, for $\gamma = -2\alpha$.

%%%%%%%%%%%%%%%%%%%%%%%%%%
\section{Conclusions}
%%%%%%%%%%%%%%%%%%%%%%%%%%

The main result of this paper is the construction of higher derivative hypermultiplet couplings to $N=(1,0)$ supergravity. The higher derivative extension of the two-derivative supergravity coupled to Yang-Mills and hypermultiplets by adding the $\alpha {\rm Riem}^2$ term and its superpartners gives rise to hypermultiplet involving higher derivative supersymmetry variations since the composite connection built out of the hyperscalars couples to all fermions. To restore supersymmetry up to first order in $\alpha$, requires addition of several new terms that involve higher derivative hypermultiplet fields. We have parametrized the most general such terms and by employing the Noether procedure we have determined the full Lagrangian and shown that only one new parameter, called $\gamma$, is needed to establish supersymmetry up to first order in the parameters $\alpha, \beta, \gamma$, where $\beta=1/g_{YM}^2$.  

Another key aspect of this construction is that we have taken the quaternionic K\"ahler space parametrized by the hyperscalars to be the quaternionic projective space $Hp(n)=Sp(n,1)/Sp(n)\times Sp(1)_R$. This is partly motivated by the fact that the dimensional reduction of Riemann-squared extension of $10D$ supergravity on $T^4$, followed by a truncation to $N=(1,0)$, yields the higher derivative coupling of four hypermultiplets  parametrizing the QK coset $SO(4,4)/SO(4)\times SO(4)$, whose truncation to single hypermultiplet yields the QK coset  $SO(4,1)/SO(4)$, which is locally the same as $Hp(1)$. This model has only the parameter that comes with the Riemann-squared invariant in $10D$, denoted by $\alpha$. Our result for the higher derivative couplings of $Hp(n)$, on the other hand, has a new independent parameter, denoted by $\gamma$. We have considered two distinct ways of truncating the coset $SO(4,4)/SO(4)\times SO(4)$ to $SO(4,1)/SO(4)$, which is locally the same as $Hp(1)$, and shown that the results agree with the truncation of our $Hp(n)$ model to $Hp(1)$, for either $\gamma=0$ or $\gamma=-2\alpha$. 

There are a number of directions to explore in view of the results of this paper. First, it would be useful to gauge the isometry group of $SO(n,1)$ or any subgroup of it thereof, as an extension of our results. In particular, it would be interesting to determine the consequences of gauging the R-symmetry group $Sp(1)_R$, or its $U(1)_R$ subgroup for our results. Next, it would be useful to establish if the higher derivative extension is possible for all symmetric QK manifolds, known as the Wolf spaces. Last, but not least, it is worth investigating possible embedding of our results in string theory which goes beyond the $Hp(1)$ model describe in this paper.

%%%%%%%%%%%%%%%%%%%%%%%%%%%%%%%%%%%%%%%%
\section*{Acknowledgements} 

We thank Guillaume Bossard, Daniel Butter, Axel Kleinschmidt and Yi Pang for useful discussions. The work of E.S. and H.C. is supported in part by  NSF grant PHY-1803875, and that of H.C. in part by the Mitchell Institute of Fundamental Physics and Astronomy.

\begin{appendix}

%%%%%%%%%%%%%%%%%%%%%%%%%%%%%%%%%%%%%%%%%%%%%%%%%
\section{Notation and Conventions}
%%%%%%%%%%%%%%%%%%%%%%%%%%%%%%%%%%%%%%%%%%%%%%%%%

To pass from the conventions of \cite{Nishino:1986dc} to those employed in this paper, we have made the redefinitions  
\be
\eta_{rs} \rightarrow - \eta_{rs}, \qquad  \gamma^r \rightarrow i \gamma^r\ , \qquad \varphi \rightarrow \frac{1}{\sqrt{2}} \varphi \ .
\ee
Thus, the spacetime signature is $\eta_{rs} ={\rm diag} (-+++++)$. For arbitrary spinors carrying un-contracted indices, we have
\be
\chi^A \gamma_{\mu_1...\mu_n} \psi^a = (-1)^{n+1} \bpsi^a \gamma_{\mu_n...\mu_1} \chi^A\ .
\ee
If the indices are the same type of symplectic indices, namely $a$ or $A$, and contracted, then an extra minus sign occurs. Thus, it follows that
\bea
&& \bpsi^a \gamma_\mu \psi^b = \bpsi^b \gamma_\mu \psi^a\ ,
\quad 
\bpsi^a \gamma_{\mu\nu\rho} \psi^b = - \bpsi^b \gamma_{\mu\nu\rho} \psi^a\ ,
\quad 
\bpsi^a \gamma_{\mu\nu\rho\lambda\tau} \psi^b = \bpsi^b \gamma_{\mu\nu\rho\lambda\tau} \psi^a\ .
\eea
Raising and lowering of symplectic indices is $\Omega^{ab} \psi_b = \psi^a$ and $\psi^a\Omega_{ab} =\psi_b$, with $\Omega_{ac}\Omega^{bc} =\delta_a^b$. 

Further definitions are:
\begin{align}
R_{\mu\nu}{}^{mn} = 2\partial_{[\mu} \omega_{\nu]}{}^{mn}  + 2\omega_{[\mu}{}^{mp} \omega_{\nu] p}{}^n \ ,
\qquad 
R = e_m{}^\mu e_n{}^\nu R_{\mu\nu}{}^{mn}\ .
\end{align}
The van der Wardeen symbols are  $(\s^a)_{A A'} = (\sigma^1, \sigma^2, \sigma^3, i )$ and $(\bar{\s}^a)^{A' A} = (\sigma^1, \sigma^2, \sigma^3,- i )$, and 
\begin{equation}
(\s^{ab})_A{}^B = (\s^{[a})_{A A'} (\bar{\s}^{b]})^{A' B} \ , \qquad (\bar{\s}^{ab})^{A'}{}_{B'} = (\bar{\s}^{[a})^{A' A} (\s^{b]})_{A B'} \ . 
\end{equation}

The following notations have been used:
\begin{align}
Q^2 &:= Q_{\m\n AB} Q^{\m\n AB}\ , & \varphi_\m &:= \partial_\mu \varphi
\nn\w2
(P^2)_{\mu\nu}^{ab} &:= P_{(\mu}^{aA} P_{\nu)}^{b}{}_A \ , & (P^2)^{ab} &:= g^{\mu\nu} (P^2)_{\mu\nu}^{ab} \ , 
\nn\w2
(P^2)_{\mu\nu} &:= (P^2)_{\mu\nu}^{ab}\,\Omega_{ba} \ , & P^2 &:= g^{\mu\nu} (P^2)_{\mu\nu}\ ,
\nn\w2
(PDP)_\mu^{AB} &:= P^{\nu a(A} D_\mu P_{\nu a}{}^{B)}\ , & (PDP)_\mu^{ab} &:= P^{\nu (a|A} D_\mu P_\nu^{b)}{}_A\ , 
\nn\w2
(PDP)_{\mu,\nu\rho} &:= P_{[\nu|}^{aA} D_\mu P_{|\rho] aA}\ .
\end{align}
Note the symmetry properties
\bea
(P^2)_{(\mu\nu)}^{ab} &=& (P^2)_{\mu\nu}^{ab} = (P^2)_{\mu\nu}^{[ab]}\ , \qquad  (P^2)^{ab} =(P^2)^{[ab]} \ , \qquad (P^2)_{\mu\nu} = (P^2)_{(\mu\nu)}\ , 
\nn\w2
(PDP)_\mu^{AB} &=& (PDP)_\mu^{(AB)}\ , \qquad (PDP)_\mu^{ab} = (PDP)_\mu^{(ab)}\ .
\eea

\section{Lowest order equations of motion}
%%%%%%%%%%%%%%%%%%%%%%%%%%%%%%%%%%%%%%%%%%%%%%%%%%%%%%%%%%%%%%

In view of the important role of the two-derivative field equations of motion that follow from the $\alpha, \beta$ and $\gamma$  independent part of the Lagrangian \eq{FL}, we record them here.  To this end we define
\bea
\E_{\m\n} &\equiv & \tfrac12 e^{-1} e^{- 2 \vp} \frac{\del \cL_0}{\del e^\m{}_a} e_{\n a} \ , \quad \E_\vp \equiv  e^{-1} e^{- 2 \vp} \frac{\del \cL_0}{\del \vp} \ , 	\quad	\E^{a A} \equiv  e^{-1} \frac{\del \cL_0}{\del \phi^\al} V^{\al a A} \ , 
\nn\w2
\E_B^{\m\n} &\equiv& 2 e^{-1} \frac{\del \cL_0}{\del B_{\m\n}} \ ,  
\quad
 \E^{\m A} \equiv \, 2 e^{-1} e^{- 2 \vp} \frac{\del \cL_0}{\del \bpsi_{\m A}} \ , \quad\E^A \equiv - \tfrac14 e^{-1} e^{- 2 \vp} \frac{\del \cL_0}{\del \bar{\chi}_A} \ ,
 \nn\w2
 \E_{a} &\equiv& - e^{-1} e^{- 2 \vp} \frac{\del \cL_0}{\del \bpsi^a} \ , 
\eea
and
\be
\qquad  {\widetilde\E}_{aA} := e^{-2\vp} \E_{aA}\ ,
\qquad \EW_{\m\n} := \E_{\m\n} + \tfrac14 g_{\m\n} \E_\varphi\ .
\ee
where the bosonic field equations are \footnote{As we shall construct the higher derivative couplings up to quartic fermion terms, we will not need the quadratic in fermion terms in the bosonic EOM's in the Noether procedure calculation.}
\begin{subequations}
\begin{align}
\mathcal{E}_\vp &=\, \tfrac12 R - 2 D_\mu \vp^\mu - 2 \vp_\mu \vp^\mu - \tfrac16 H^2 - P^2 \ ,
\label{b1}
\w2
\mathcal{E}_{\m\n} &=\, \tfrac14 R_{\m\n} - \tfrac12 \vp_{\mu\nu} - \tfrac14 H^2_{\m\n} - \tfrac12 (P^2)_{\m\n} - \tfrac14 \mathcal{E}_\varphi g_{\m\n}\ ,
\label{b2}
\w2
{\cal E}_{aA} &= D_\mu \left(e^{2\varphi} P^\mu_{aA}\right)\ ,
\label{b3}
\w2
\E_B^{\mu\nu} &= D_\rho\big(e^{2\varphi} H^{\mu\nu\rho}\big)\ ,
\label{b4}
\end{align}
\label{beom}
\end{subequations}
and the fermionic field equations are
\begin{subequations}
\begin{align}
\E^{\mu A} 
&= \gamma^{\mu\nu\rho} \psi_{\nu\rho}^A +4 \gamma^{\m\n} D_\n \chi^A + \tfrac16 \gamma^{[\mu} \gamma\cdot H \gamma^{\nu]} \psi_\nu^A +\tfrac13 \gamma^{\m\n\rh\s} H_{\n\rh\s} \chi^A + 2\vp^\mu \gamma^\nu \psi_\nu^A
\nn\\
&
+\Big( 2 \gamma^{\m\n\rh} \psi_\rh^A -2\gamma^\mu \psi^{\nu A} + 8\gamma^{\m\n} \chi^A +4\gamma^\nu\gamma^\mu\chi^A \Big)\vp_\nu-2\gamma^\nu\gamma^\mu \psi_a P_\nu^{aA}\ ,
\label{f1}
\w2
\E^A 
&= \slashed{D} \chi^A +\tfrac14 \gamma^{\m\n} \psi_{\m\n}^A +\tfrac{1}{24} \gamma^{\m\n\rh\s} \psi_\s^A H_{\m\n\rh} +\tfrac{1}{12}\gamma\cdot H \chi^A-\tfrac12\gamma^\nu\gamma^\mu \psi_\nu^A \vp_\mu 
\nn\\
&
+\tfrac12 \gamma^\mu\psi_a P_\mu^{aA} +\gamma^\mu \chi^A \vp_\mu\ ,
\label{f2}
\w2
\E_a &= \slashed{D} \psi_a  + \big( -\gamma^\nu \gamma^\mu \psi_\nu^A  +2 \gamma^\mu \chi^A \big) P_{\mu aA}  +\gamma^\mu \psi_a \vp_\mu  +\tfrac{1}{12} \gamma\cdot H \psi_a\ . 
\label{f3}
\end{align}
\label{feom}
\end{subequations}
where we have introduced the notation $\vp_{\m\n} := D_\m \pd_\n \vp$, and it is important to note that in the fermionic field equations above, $\psi_{\m\n}= \psi_{\m\n}(\omega)= 2D_{[\m} \psi_{\n]}$, unless stated otherwise.  It follows for the EOM's given above that
\begin{subequations}
\begin{align}
R_{\m\n} &=  4 \EW_{\m\n}   +2 \vp_{\mu\nu} + H^2_{\m\n} +2 (P^2)_{\m\n}\ ,
\w2
D_\m \vp^\m &= 2 \E_\vp + 2 \E^\m{}_\m  - 2 \vp^\m \vp_\m + \tfrac13 H^2 \ ,
\w2
D_\m P^\m_{aA} &=   \EW_{aA} - 2\vp_\m P^\m_{aA}\ ,
\label{be3}
\w2
D_\rho H^{\mu\nu\rho} &= e^{-2\vp} \E_B^{\mu\nu} -2\vp_\rh H^{\rh\m\n}\ ,
\end{align}
\label{be}
\end{subequations}
and 
\begin{subequations}
\begin{align}
\gamma^\la \psi_{\la\mu}^A &= \tfrac12 \E_\mu^A -2 \gamma_\mu \E^A  + 2D_\mu \chi^A
+2P_\mu^{aA} \psi_a - \vp_\nu \gamma^\nu\psi_\mu^A 
 \nn\\
 & -\tfrac{1}{12} \gamma\cdot H \psi_\mu^A  +\tfrac14 H_{\mu\nu\rho} \Big(\gamma^{\nu\rho\sigma} \psi_\sigma^A
-2\gamma^\nu \psi^{\rho A} + 2\gamma^{\nu\rho} \chi^A  \Big)\ ,
\label{fe1}
\w2
\slashed{D} \chi^A &= \tfrac14 \gamma^\mu \E_\mu^A-4\E^A  -\tfrac12 \big(\gamma^{\mu\nu} \psi_\nu^A  -\psi^{\mu A} \big) \vp_\mu -2\gamma^\mu \chi^A \vp_\mu 
\nn\\
& +\tfrac{1}{12} H_{\mu\nu\rho} \big( \gamma^{\m\n\rh\s} \psi_\s^A -3\gamma^{\mu\nu} \psi^{\rh A} +\gamma^{\m\n\rh} \chi^A \big)\ ,
\w2
\slashed{D} \psi_a &= \E_a  + \big( \gamma^\nu \gamma^\mu \psi_\nu^A  -2 \gamma^\mu \chi^A \big) P_{\mu aA}  -\gamma^\mu \psi_a \vp_\mu  -\tfrac{1}{12} \gamma\cdot H \psi_a\ . 
\label{cfe}
\end{align}
\label{fe}
\end{subequations}
In the lowest order EOMs given above, it is understood that $H=dB$.

We will also need the following relations which follow from differentiation of \eq{fe1},
\begin{align}
\slashed{D} \psi_{\n\rh}^A \Big|_{[\n\rh]} 
&= \tfrac14 R_{\n\rh\la\ta} \Big( \g^\m\g^{\la\ta} \psi_\m^A   + 2\g^{\la\ta} \chi^A \Big)
- 4 ( D_\n \psi^a ) P_{\rh a}{}^A -2 \g^\ta \psi_\n^A (P^2)_{\rh\ta} 
\nn\w1
& -Q_{\n\rh}{}^A{}_B \Big(  \g^\m \psi_\m^B +2 \chi^B \Big)
 -2\g^\m \psi_\rh^B Q_{\m\n}{}^A{}_B  - \g^\ta \psi^A_{\n\rh} \vp_\ta - \g^\ta \psi^A_\n H^2_{\rh\ta} 
\nn\w2
& + D_\n \Big( \g_\ta \psi^A_\la H_\rh{}^{\la\ta} + \tfrac12 \g_{\s\la\ta} \psi^{\s A} H_\rh{}^{\la\ta} - \tfrac16 \g_{\s\la\ta} \psi^A_\rh H^{\s\la\ta} + \g_{\la\ta} \chi^A H_\rh{}^{\la\ta} \Big)
\nn\w2
&  -4 \g^\ta \psi_\n^A \EW_{\rh\ta} + D_\n \Big( \mathcal{E}_\rh^A - 4 \g_\rh\mathcal{E}^A \Big) \ ,
\label{id1}
\w2
D^\m \psi_{\m\n}^A &= \tfrac14 \g^{\rh\s} \psi^{\m A} R_{\m\n\rh\s} + \tfrac12 \g^{\rh\s} \psi_\n^B Q_{\rh\s}{}^A{}_B - \psi^{\m B} Q_{\m\n}{}^A{}_B 
\nn\w2
&\, + \tfrac12 \psi_\n^A P^2 + \g^\m ( D_\n \psi^a ) P_{\m a}{}^A - \g^\m \psi^a D_\m P_{\n a}{}^A 
\nn\w2
&\, - 2 \g_\m ( D_\n \chi^A ) \vp^\m- \psi^A_{\m\n} \vp^\m - \g^{\m\rh} \psi^A_{\m\n} \vp_\rh + 2 \g^\m \psi^a P_{\n a}{}^A - \tfrac12 \psi^A_\n \vp^\m{}_\m 
\nn\w2
&\, + \tfrac14 \psi^A_\n H^2 - \tfrac12 \psi^{\m A} H^2_{\m\n} - \tfrac12 \g^{\rh\s} \psi^A_\rh H^2_{\s\n} + \g^\m \chi^A H^2_{\m\n} + \tfrac16 \g^{\rh\s\la} \psi^a P_{\n a}{}^A H_{\rh\s\la} 
\nn\w2
&\, - \tfrac1{12} D^\m \Big( \g_\m \g_{[\n|} \g_{\rh\s\la} \g_{|\ta]} \psi^{\ta A} H^{\rh\s\la} - \g_{\m\n} \g_{\rh\s\la\ta} \psi^{\ta A} H^{\rh\s\la} \Big) 
\nn\w1
&\, + D^\m \Big( - \g_\rh \chi^A H_{\m\n}{}^\rh + \tfrac12 \g_{\m\s\la} \chi^A H_\n{}^{\s\la} \Big) - \tfrac16 D_\n \Big( \g_{\rh\s\la} \chi^A H^{\rh\s\la} \Big) 
\nn\w2
&\, + \psi_\n^A \EW^\m{}_\m - 2 \psi^{\m A} \EW_{\m\n} - 2 \g^{\rh\s} \psi_\rh^A \EW_{\s\n} 
\nn\w2
&\, + 4 \g^\m \chi^A \EW_{\m\n} - 2 \E^a P_{\n a}{}^A + \tfrac12 \slashed{D} \E_\n^A  - 2 \g_{\m\n} D^\m \E^A \ .
\label{id2}
\end{align}
%

%%%%%%%%%%%%%%%%%%%%%%%%%%%%%%%%%%%%%%%%%%%%%%%%%%%%%%%%%%%%%%
\section{Identities not involving the equations of motion}
%%%%%%%%%%%%%%%%%%%%%%%%%%%%%%%%%%%%%%%%%%%%%%%%%%%%%%%%%%%%%%

In what follows we list lemmas which have been used in simplifying the variation of the Lagrangian. 
\allowdisplaybreaks{
\begin{align}
P_\nu{}^b{}_A (P^2)^{\mu\nu}_{ab} &= - \tfrac14 P_{\nu a}{}^B Q^{\mu\nu}_{AB} + \tfrac14 P_{\nu aA} (P^2)^{\mu\nu} + \tfrac14 P^\mu_{aA} P^2
\w2
P^{\m b}{_A} (P^2)^{\n\rh}_{ab}\Big|_{[\mu\nu]} &= \tfrac14 P^\rho{}_a{}^B Q^{\mu\nu}_{AB} - \tfrac14 P^\mu{}_a{}^B Q^{\nu\rho}_{AB} + \tfrac14 (P^2)^{\mu\rho} P^\nu_{aA}
\w2
P_\mu{}^b{}_A (PDP)^\mu_{ab} &= \tfrac12 P_{\mu a}{}^B (PDP)^\mu_{AB} - \tfrac14 (P^2)_{\mu\nu} D^\mu P^\nu_{aA} + \tfrac18 P^\mu_{aA} \partial_\mu P^2 
\w2
(PDP)^\mu_{ab}  P^{\nu b}{}_A \Big|_{[\mu\nu]} &= - \tfrac18 P^\lambda{}_a{}^B D_\lambda Q^{\mu\nu}{}_{AB} + \tfrac14 Q^{\mu\lambda}{}_{AB} D^\nu P_{\lambda a}{}^B + \tfrac14 (P^2)^{\mu\lambda} D^\nu P_{\lambda aA}
\nn\w2
& - \tfrac14 P_{\la aA} (PDP)^{\la,\m\n}
\w2
(P^2)^{ab} (P^2)_{ab} &= \tfrac14 Q^2 + \tfrac12 (P^2)_{\mu\nu} (P^2)^{\mu\nu}\ ,
\w2
Q_{\m\n}^{ab} Q^{\m\n}_{ab} &= \tfrac12 Q^2 + (P^2)^2 -  (P^2)_{\m\n} (P^2)^{\m\n}
\label{C5}\w2
D_{[\mu} (P^2)_{\nu]\rho} &= -(PDP)_{\rho,\mu\nu}\ ,
\w2
D_{[\m} (PDP)_{\n] AB} &= \big( D_{[\m|} P^{\rh a}{}_{(A|} \big) \big( D_{|\n]} P_{\rh a |B)} \big) + \tfrac14 R_{\m\n}{}^{\rh\s} Q_{\rh\s AB} 
\nn\\
&\, - \tfrac14 ( Q_{[\m}{}^\rh Q_{\n] \rh} )_{AB} - \tfrac12 (P^2)_{[\m}{}^\rh Q_{\n] \rh AB} + \tfrac14 Q_{\m\n AB} P^2 
\w2
P_\rho{}^a{}_A D_\mu P_{\s\, a B} \Big|_{[\rho\sigma]} &= \tfrac14 D_\mu Q_{\rho\sigma AB} - \tfrac12 \epsilon_{AB}  (PDP)_{\mu,\rho\sigma}
\w2
P^{\nu aA} D_\mu P_{\nu}{}^b{}_A &= (PDP)_\mu^{ab} +\tfrac12 D_\mu (P^2)^{ab}
\w2
D_\mu D_\nu P^2 &= \, 2 \big( D_\m P^{\rh a A} \big) \big( D_\n P_{\rh a A} \big) + 2 P^{\rh a A} D_\rh D_\m P_{\n a A} + 2 (P^2)^{\rh\s} R_{\m\rh\n\s} 
\nn\\
& + \tfrac32 \big( Q_\m{}^\rh Q_{\n\rh} \big) + (P^2)_{\m\n} P^2 - (P^2)_\m{}^\rh (P^2)_{\n\rh} 
\label{ddp}
\w2
P^{\m b}{}_A D^\n (P^2)_{ab} \Big|_{[\mu\nu]} &= \tfrac14 P^\lambda{}_a{}^B D_\lambda Q^{\mu\nu}{}_{AB} + \tfrac12 Q^{\mu\lambda}{}_{AB} D^\nu P_{\lambda a}{}^B + \tfrac12 (P^2)^{\mu\lambda} D^\nu P_{\lambda aA} 
\nn\w2
& + \tfrac12 P_{\rh\, aA} (PDP)^{\rh,\m\n} 
 \w2
Q_{\mu\rho a}{}^b D_\nu P^\rho_{bA} \Big|_{[\mu\nu]} &= P_{\mu a}{}^B (PDP)_{\nu AB} - \tfrac14 P_{\mu aA} \partial_\nu P^2 - \tfrac14 P_{\lambda a}{}^B D_\lambda Q_{\mu\nu AB} + \tfrac12 P^\lambda_{aA} (PDP)_{\lambda,\mu\nu} 
\w2
P^{\m b}{}_A D_\m (P^2)_{ab} &=  P_{\mu a}{}^B (PDP)^\mu_{AB} +\tfrac12 (P^2)_{\mu\nu} D^\mu P^\nu_{aA} +\tfrac14 P^\mu_{aA} \partial_\mu P^2
\end{align}
}

%%%%%%%%%%%%%%%%%%%%%%%%%%%%%%%%%%%%%%%%%%%%%%%%%%%%%%%%%
\section{Identities involving equations of motions}
%%%%%%%%%%%%%%%%%%%%%%%%%%%%%%%%%%%%%%%%%%%%%%%%%%%%%%%%%
The following relations hold modulo  the $\vp_\m$ and $H$ dependent terms.
\begin{align}
D^\m Q_{\m\n}^{AB} &=\, 2 (PDP)_\n^{AB} - 2 P_\nu^{a(A} \EW_a{}^{B)} 
\w2
D^\la (P^2)_{\la\m} &= \tfrac12 \pd_\m P^2 + \EW^{aA} P_{\m aA} 
\w2
D^\mu Q_{\mu\nu}^{ab} &= 2 (PDP)_\nu^{ab} - 2 \EW_A^{(a} P_\nu^{b)A} 
\w2
D^\rh (PDP)_{\m, \n\rh} &= \tfrac12 \big(D_\m P^{\rh aA}\big) \big(D_\n P_{\rh aA}\big) - \tfrac12 P^{\rh\, aA} D_\rh D_\m P_{\n aA} 
\label{dp12}
\nn\\
& + \tfrac38 \big(Q_{\m\rh} Q_\n{}^\rh \big) + \tfrac34 (P^2)_{\m \rh} (P^2)_\n{}^\rh + \tfrac14 (P^2)_{\m\n} P^2 
\nn\\
& + \tfrac12 P_\n{}^{aA} D_\m \EW_{aA} - \tfrac12 \EW^{aA} D_\m P_{\n aA} + 2 \EW_{\m\rh} (P^2)^\rh{}_\nu 
\w2
D_\m (PDP)^\m_{AB} &= ( D_\mu \EW_{a(A} ) P^{\mu a}{}_{B)} 
\w2
P_\mu{}^b{}_A D_\nu (P^2)^{\mu\nu}_{ab} &= \tfrac14 (P^2)_{\mu\nu} D^\mu P^\nu_{aA} +\tfrac12 P_{\nu a}{}^B (PDP)^\nu_{AB} + \tfrac18 P^\mu_{aA} \partial_\mu P^2 +\tfrac14 P^2 \EW_{aA}
\nn\w2
& +\tfrac12 P^\m{}_{b (A|} P_{\m a |B)} \EW^{bB} -\tfrac14 (P^2)_{ab}\,\EW^b{}_A 
\w2
P_{\n a}{}^B D_\m Q^{\m\n}{}_{AB} &= 2 P_{\mu a}{}^B (PDP)^\mu_{AB} + \tfrac32 (P^2)_{ab}\,\EW^b{}_A - P^\m{}_{a (A|} P_{\m b |B)} \EW^{bB} 
\label{5id1}
\w2
P_{\nu aA}D_\mu (P^2)^{\mu\nu} &= \tfrac12 P_{\mu a A} \partial^\mu P^2 + \tfrac12 (P^2)_{ab}\,\EW^b{}_A + P^\m{}_{a (A|} P_{\m b |B)} \EW^{bB}
\w2
P_\m{^b}{_A} D_\n Q^{\m\n}{}_{ab} &= - P_{\m a}{}^B (PDP)^\m_{AB} + \tfrac12 (P^2)^{\m\n} D_\m P_{\n\,aA} - \tfrac14 P^\m_{aA} \partial_\m P^2 - \tfrac12 P^2 \EW_{aA} 
\nn\w2
& + P^\m{}_{b A} P_{\m a B} \EW^{bB} 
\w2
P^\m{}_a{}^B D_\rh Q^{\rh\n}{}_{AB} \Big|_{[\mu\nu]} &= 2 P^\mu{}_a{}^B (PDP)^\nu_{AB} +\tfrac34 Q^{\mu\nu}_{ab}\,\EW^b{}_A - P^\mu_{[a|A} P^\nu_{|b]B}\,\EW^{bB} 
\w2
P^\m_{aA}  D_\rh (P^2)^{\rh\n} \Big|_{[\mu\nu]} &=
\tfrac12 P^\m_{aA} \partial^\n P^2 + \tfrac14 Q^{\mu\nu}{}_{ab}\,\EW^b{}_A + P^\m_{[a|A} P^\n_{|b]B}\,\EW^{bB} 
\w2
P^{\m b}{}_A D_\rh (P^2)^{\n\rh}_{ab}\Big|_{[\m\n]} &= \tfrac18 P^\rh{}_a{}^B D_\rh Q^{\mu\nu}{}_{AB} + \tfrac14 Q^{\m\rh}{}_{AB} D^\nu P_{\rh a}{}^B 
\nn\w2
& + \tfrac14 (P^2)^{\m\rh} D^\nu P_{\rh aA} + \tfrac14 P_{\rh\, aA} (PDP)^{\rh,\m\n}
\nn\w2
& + \tfrac14 Q^{\mu\nu}{}_{AB}\,\EW_a{}^B  + \tfrac18Q^{\mu\nu}_{ab}\EW^b{}_A - \tfrac12 P^\m_{[a|A} P^\n_{|b]B}\,\EW^{bB} 
\w2
\Box P^\m_{aA} &= \tfrac32 (P^2)^{\m\n} P_{\n aA} +\tfrac32 Q^{\m\n}{}_{AB} P_{\n a}{}^B + \tfrac12 P^\m{}_{aA} P^2 + D^\mu \EW_{aA} + 4 \EW^{\m\n} P_{\n aA}\ ,
\w2
\Box Q_{\m\n AB} &=\, \Big[ 4 \big( D_\m P^{\rh a}{}_A \big) \big( D_\n P_{\rh a B} \big) + 3 \big( Q_\m{}^\rh Q_{\n\rh} \big)_{(AB)} - 6 (P^2)_\m{}^\rh Q_{\n\rh AB} 
\nn\\
& + Q_{\m\n AB} P^2 - 4 P_{\m a}{}_{(A} D_\n \EW^a{}_{B)} + 8 Q_\m{}^\rh{}_{AB} \EW_{\n\rh} \Big]\Big|_{[\m\n]}
\w2
\Box P^2 &= \tfrac32 Q^2 + 3 (P^2)^{\m\n} (P^2)_{\m\n} + (P^2)^2 + 2 \big( D^\m P^{\n a A} \big) \big( D_\m P_{\n a A} \big)
\nn\w2
& + 2 P_\m{}^{aA} D^\m \EW_{aA} + 8 (P^2)^{\m\n} \EW_{\m\n} \ . 
\label{bp}
\end{align}
The $\vp_\m$ and $H$-dependent terms can simply be obtain in all teh equations by by letting
\begin{align}
\EW_{a A} \to &\, \EW_{a A} - 2 P^\m{}_{a A} \vp_\m 
\nn\w2
\EW_{\m\n} \to &\, \EW_{\m\n} + \tfrac12 \vp_{\m\n} + \tfrac14 H^2_{\m\n} \ .
\end{align}
\label{rr}

\end{appendix}

\end{document}